\documentclass[aps,groupedaddress,superscriptaddress,amsmath,amssymb,twocolumn,prl]{revtex4-2}

\usepackage[utf8]{inputenc}
\DeclareUnicodeCharacter{0327}{\c}
\usepackage{amssymb}
\usepackage{amsmath}
\usepackage{bm}
\usepackage{graphicx}
\usepackage{varioref}
\usepackage{braket}
\usepackage{siunitx}
\usepackage[normalem]{ulem}

\usepackage{setspace}
\usepackage{textcomp}

\usepackage[colorlinks=true,citecolor=blue,linkcolor=blue,urlcolor=blue, linktocpage=true,breaklinks=true,pagebackref=false]{hyperref}

\begin{document}

\title{Non-Exponential Relaxation in the Rotating Frame of a Driven Nanomechanical Mode}

\author{Hyunjin Choi}
\affiliation{Department of Physics, Korea Advanced Institute of Science and Technology (KAIST), Daejeon, 34141, Republic of Korea}

\author{Oriel Shoshani}
\thanks{Corresponding author: oriels@bgu.ac.il}
\affiliation{Department of Mechanical Engineering, Ben-Gurion University of the Negev, Be'er-Sheva 84105, Israel}

\author{Ryundon Kim}
\author{Younghun Ryu}
\thanks{Present address: Institute of Physics and Center for Quantum Science and Engineering, École Polytéchnique Fédérale de Lausanne (EPFL), Lausanne 1015, Switzerland}
\author{Jinhoon Jeong}
\affiliation{Department of Physics, Korea Advanced Institute of Science and Technology (KAIST), Daejeon, 34141, Republic of Korea}

\author{Junho Suh}
\affiliation{Department of Physics, Pohang University of Science and Technology (POSTECH), Pohang, 37673, Republic of Korea}

\author{Steven W. Shaw}
\affiliation{Department of Mechanical and Civil Engineering, Florida Institute of Technology, Melbourne, Florida 32901, USA}
\affiliation{Department of Mechanical Engineering, Michigan State University, East Lansing, Michigan 48824, USA}

\author{M. I. Dykman}
\affiliation{Department of Physics and Astronomy, Michigan State University, East Lansing, Michigan 48824, USA}

\author{Hyoungsoon Choi}
\thanks{Corresponding author:  h.choi@kaist.ac.kr}

\affiliation{Department of Physics, Korea Advanced Institute of Science and Technology (KAIST), Daejeon, 34141, Republic of Korea}
\affiliation{Graduate School of Quantum Science and Technology, Korea Advanced Institute of Science and Technology (KAIST), Daejeon, 34141, Republic of Korea}

\begin{abstract}

We present direct observation of the ring-down dynamics in the rotating frame of a resonantly driven single-mode nonlinear nanomechanical resonator. An additional close to resonance harmonic force excites nonlinear oscillations about the fixed point in the rotating frame. When the secondary drive is removed,  we measure decay  of the in-phase and quadrature components toward this fixed point. We show that the decay  of the in-phase signal is non-exponential, even though the vibration amplitude decays exponentially if both forces are switched off. A minimalistic model  captures these dynamics as well as the spectrum of the vibrations excited by the additional force, relating them to the dissipation-induced symmetry breaking of the dynamics in the rotating frame.

\end{abstract}

\maketitle

Decay of vibrations of nanomechanical systems reveals important information about the physics of the vibrational modes, such as their linear or nonlinear coupling to other excitations, including the mode-mode coupling. The decay rate imposes the limits on various applications of the mechanical modes, in particular those in classical and quantum sensing. \cite{Ekinci2005, Unterreithmeier2010, chen2017, Clerk2010,bachtold2022}. Measuring the spectral linewidth is usually sufficient to determine the decay rate of many vibrational systems, especially at macro- and micro-scales. However, in nanomechanical systems, spectral linewidths can be broadened by frequency fluctuations, obscuring the true decay rate \cite{Maillet2017, Zhang2015, Fong2012, Huang2019}. To overcome this limitation, ring-down measurements are widely adopted \cite{schmid2011, Güttinger2017, Wang2022, chen2017, Shoshani2017}.

In a ring-down measurement, the system is actively driven initially, then the drive is turned off, and the resulting change in energy or amplitude over time is directly tracked. This transient response allows one to separate out dissipation from dephasing effects \cite{Schneider2014}. A major dissipation mechanism  in nanomechanical systems is the coupling to a thermal reservoir, which is linear in the mode coordinate; it leads to  a rather simple exponential decay of the amplitude \cite{bachtold2022}. However, the decay may be more complicated, and ring-down measurements can  detect such  interesting characteristics of the systems as nonlinear (amplitude-dependent) damping, nonlinear frequency shifts, and mode coupling \cite{Eichler2011,Shoshani2017,chen2017,Polunin2016}.

The rotating frame offers a simple view of intricate nonlinear dynamics, and therefore, it is widely used when analyzing nanomechanical oscillators. In this frame, a periodic response to a resonant drive is described by a fixed point. Studying vibrational dynamics in the rotating frame has unearthed a wealth of phenomena, including fluctuation squeezing \cite{Huber2020, Ochs2021, Yang2021}, the onset of spectral sidebands \cite{Zhang2024, Houri2021}, frequency combs \cite{Czaplewski2018,Dykman2019,Ochs2022,keskekler2022, Jong2023, Wang2022comb, Singh2020, Mouharrar2024}, the occurrence of chaos \cite{Houri2020, Defoort2021}, and manifestations of mode coupling \cite{Shoshani2023, Eriksson2023, Houri2019, Luo2021, Czaplewski2019}. However, the relaxation from an excited state to the fixed point in the rotating frame has remained largely unexplored. 

In this paper, we tackle the problem of decay directly by studying the ring-down of a nanomechanical system in the rotating frame. We find that a driven nonlinear nanomechanical oscillator with a very small damping exhibits an unexpectedly strong nonlinear damping in the rotating frame, accentuated by a very peculiar non-exponential relaxation. The effect is pronounced even  where the nonlinear part of the vibration energy in the laboratory frame remains small compared to the linear part.

Our system is subjected to a near-resonant harmonic drive in the lab frame, which can be used to tune the dynamics in the rotating frame and its corresponding fixed point. The harmonic drive also breaks the time-translational symmetry of the system and provides a reference point in the rotating frame to phase-lock the response to a weak secondary drive that we apply. The secondary drive is used to perturb the system into a state of forced periodic vibrations about the fixed point, which manifests as a frequency comb in the lab frame \cite{Dykman2019, Ochs2022}.

By turning off the secondary drive, we observe the phase-locked ring-down, {\it i.e.}, relaxation towards the fixed point as the energy is dissipated. The ring-down response shows different decay patterns for the in-phase and quadrature components owing to strong nonlinearities present only in the rotating frame. The distinct decay rates observed for the in-phase and quadrature components are ultimately related to the lack of symmetry with respect to time translation by a quarter of the modulation period, which would lead to the interchange of the components. In addition, the system lacks  inversion symmetry in the rotating frame, which results in significant contributions of even harmonics to the frequency content of the dynamical variables -- a clear signature of the underlying strong nonlinearities.

\begin{figure}[t]
\centering
\includegraphics[width=1\linewidth]{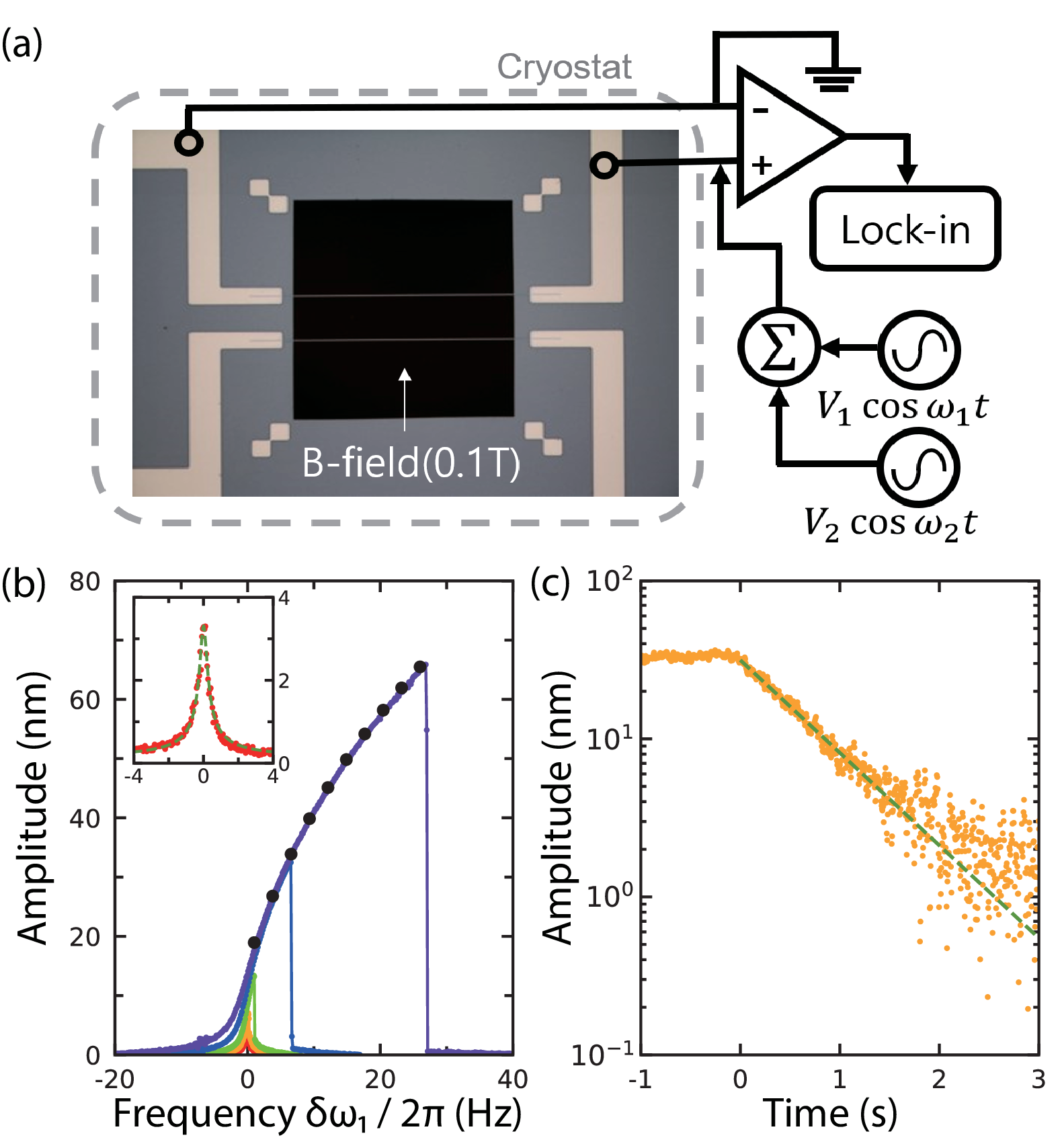}
\caption{ (a) Optical image of the resonator and measurement setup. The experiment is conducted on one of the two suspended strings. (b) The nonlinear resonance curves in the lab frame of the first flexural mode were obtained from upward frequency sweeps. The rainbow-colored curves represent measured amplitude spectra at increasing drive strengths. The black dots represent the theoretical approximation values along the curve. The inset shows a resonance curve in the linear regime with drive $F_{1}=\qty{0.015}{\meter / \second^2}$, and the green dashed curve represents a fit to the amplitude response of a driven damped harmonic oscillator. (c) Ring-down response of the first flexural mode amplitude in the Duffing nonlinear regime under a drive of $F_{1}=\qty{0.15}{\meter / \second^2}$ and $\delta \omega_1/2\pi=\qty{6.5}{\hertz}$. The green dashed line represents the fitted linear decay.}
\label{fig:1}
\end{figure}

The system we study is a doubly clamped beam in a tensile-stress-dominated regime \cite{Bückle2021, Klaß2022}. The device is a heterostructure composed of {\qty{100}{nm}} thick silicon nitride and a {\qty{50}{nm}} thick aluminum layer deposited on top for carrying an alternating current (AC). The beam is $L= \qty{360}{\micro\meter}$ long and $w=\qty{1}{\micro\meter}$ wide as shown in Fig.~\ref{fig:1}(a). The bare mass of the wire from its physical dimension is {$m = \qty{ 1.63d-13 }{\kilo\gram}$}. The device is housed in an experimental cell and cooled to approximately $T\approx4K$. The low-vacuum environment ($\sim10^{-6}$ mbar) ensures low-damping conditions.

We focus on the fundamental flexural mode of the nanoresonator. Its dynamics is well described by the following minimalistic model
\begin{equation}
\ddot{x} + 2\Gamma\dot{x} + \omega_{0}^{2}x + \gamma x^{3} = F_{1}\cos(\omega_{1}t) + F_{2}\cos(\omega_{2}t),
\label{eq:gov_eq}
\end{equation}
where $x$ is the modal displacement, $\Gamma$, $\omega_0$, and $\gamma$ are the damping, natural frequency, and Duffing coefficients of the mode, respectively, and we use units where  the mode mass is set equal to 1. The primary and secondary forces on the right-hand side of Eq. \eqref{eq:gov_eq} have drive amplitudes $F_{1}$ and $F_{2}$, and drive frequencies $\omega_1$ and $\omega_2$, respectively. While the driving frequencies, $\omega_1$ and $\omega_2$, are kept close to the mode eigenfrequency, the magnitude of the forces differs significantly, with $F_1 \gg F_2$, and hence $F_1$ is designated as the primary drive.

The measured  eigenfrequency of the mode is $\omega_{0}/2\pi \approx \qty{248.013}{\kilo \hertz}$ and the relaxation rate is $\Gamma = \qty{1.35}{\hertz}$ ($Q\approx580\,000$). (See the inset of Fig.~\ref{fig:1}(b) and \ref{fig:1}(c).) The characteristic exponential decay of this mode, observed in the Duffing nonlinear regime in the lab frame, is confirmed by the linear nature of the amplitude decay, which has slope $\Gamma$ on a log-linear plot depicted in Fig.~\ref{fig:1}(c). Fig.~\ref{fig:1}(b) illustrates a Duffing stiffening response. The Duffing nonlinearity is quantified from the backbone curve; using $\omega_{\rm max}-\omega_0=3\gamma a_{\rm max}^2/(8\omega_0)$, where $\omega_{\rm max}$ and $a_{\rm max}$ are the maximal values of frequency and amplitude reached on the response curve, Fig.~\ref{fig:1}(b), we obtain $\gamma = \qty{1.597d23}{\meter^{-2} \second^{-2}}$ With the experimentally evaluated $\Gamma$, $\omega_0$, and $\gamma$, we find the following critical drive level for the onset of bistability $F_{\rm cr_1}=16\sqrt{\omega_0^3\Gamma^3/(9\sqrt{3}\gamma)}=\qty{0.033}{\meter/\second^2}$ \cite{landau1976}.

Our primary interest lies in the rotating frame dynamics associated with the large-amplitude branch of the Duffing oscillator, as described by Eq. \eqref{eq:gov_eq}, in the regime where $F_{2} \ll F_{1}$ and $|\omega_{2}-\omega_{1}|/\omega_{0} \ll 1$. We change variables to $q$, $p$ using $x(t)=q(t)\cos(\omega_{1}t)+p(t)\sin(\omega_{1}t)$ and the corresponding equation for $\dot x$. Then we obtain, in the rotating wave approximation (RWA), the following Hamiltonian $\mathcal{H}=\mathcal{H}_0(q,p)+\mathcal{H}_1(q,p,t)$ where 
\begin{align*}
&\mathcal{H}_0(q,p)=\frac{3\gamma}{32\omega_1}(q^{2}\!+\!p^{2})^2 - \frac{1}{2}\delta \omega_1(q^2\!+\!p^2) -\frac{F_{1}q}{2\omega_{1}},\\ 
&\mathcal{H}_1(q,p,t)= - {F_{2}}[q\cos(\delta\omega_2 t)-p\sin(\delta\omega_2 t)]/2\omega_1
\end{align*}
and the frequency detunings are given by $\delta \omega_{1} = \omega_{1} - \omega_{0}$, $\delta \omega_{2} = \omega_{2} - \omega_{1}$. The second frequency detuning, $\delta\omega_2$, is defined relative to $\omega_1$ rather than $\omega_0$, as the rotating frame is chosen at frequency $\omega_1$. From Eq. \eqref{eq:gov_eq}, we obtain the RWA-equations of motion for $q,p$ of the form $\dot{q}=\partial_{p} \mathcal{H} -\Gamma q, \; \dot{p}=-\partial_{q}\mathcal{H}-\Gamma p$. 

When $F_2=0$, and if we disregard damping, the mode performs vibrations  in the rotating frame about the fixed point $(q_\mathrm{eq}, p=0)$, with   $q_{\rm eq}$  being the larger solution of the equation $F_1/(2\omega_1)+\delta\omega_1q_{\rm eq}-3\gamma q_{\rm eq}^2/(8\omega_1)=0$. Where the amplitude of these vibrations is small, their frequency is given by $\Omega=\sqrt{\omega^{(1)}\omega^{(2)}}$, with $\omega^{(1)}=\partial_p^2\mathcal{H}|_{q_{\rm eq}}=\left(3\gamma q^2_{\rm eq}/(8\omega_1)-\delta\omega_1\right)$, $\omega^{(2)}=\partial_q^2\mathcal{H}|_{q_{\rm eq}}=\left(9\gamma q^2_{\rm eq}/(8\omega_1)-\delta\omega_1\right)$. The weak damping of our system allows us to consider the regime where $\Gamma$ is small not only compared to $\omega_0$, but also compared to $\Omega$, i.e., the vibrations are underdamped in the rotating frame.

In the absence of damping ($\Gamma=0$) and secondary drive ($F_2=0$), the equations of motion $\dot q=\partial_p\mathcal{H}_0$ and $\dot p=-\partial_q\mathcal{H}_0$ can be rewritten in terms  of the action of the harmonic oscillator $I_{\rm ho}\equiv(q^2+p^2)/2$ (cf. \cite{landau1976})
\begin{align}
q=\frac{3\gamma}{4F_1}I_{\rm ho}^2-\frac{2\omega_1\delta\omega_1}{F_1}I_{\rm ho}-\frac{2\omega_1\mathcal{H}_0}{F_1},~p=\frac{2\omega_1}{F_1}\dot I_{\rm ho},
\label{eq:q&p_con}
\end{align}
The time evolution of $I_{\rm ho}$ is determined by  the first-order equation $ \dot I_{\rm ho}^2/2+U_{\rm eff}(I_{\rm ho})=0$ with an effective potential $U_\mathrm{eff}$ and can be described explicitly in terms of the Jacobi elliptic functions, see SM.  It describes in the explicit form periodic oscillations of $I_\mathrm{ho}$, and thus of the dynamical variables $q,p$,  with frequency that depends on the value of $\mathcal{H}_0$. This solution allows us to analyze,  in the rotating frame, the linear and nonlinear response of the mode to the secondary drive. In particular, it enables studying the analog of the ``backbone'' curve of such response.

\begin{figure}[t]
\centering
\includegraphics[width=1\linewidth]{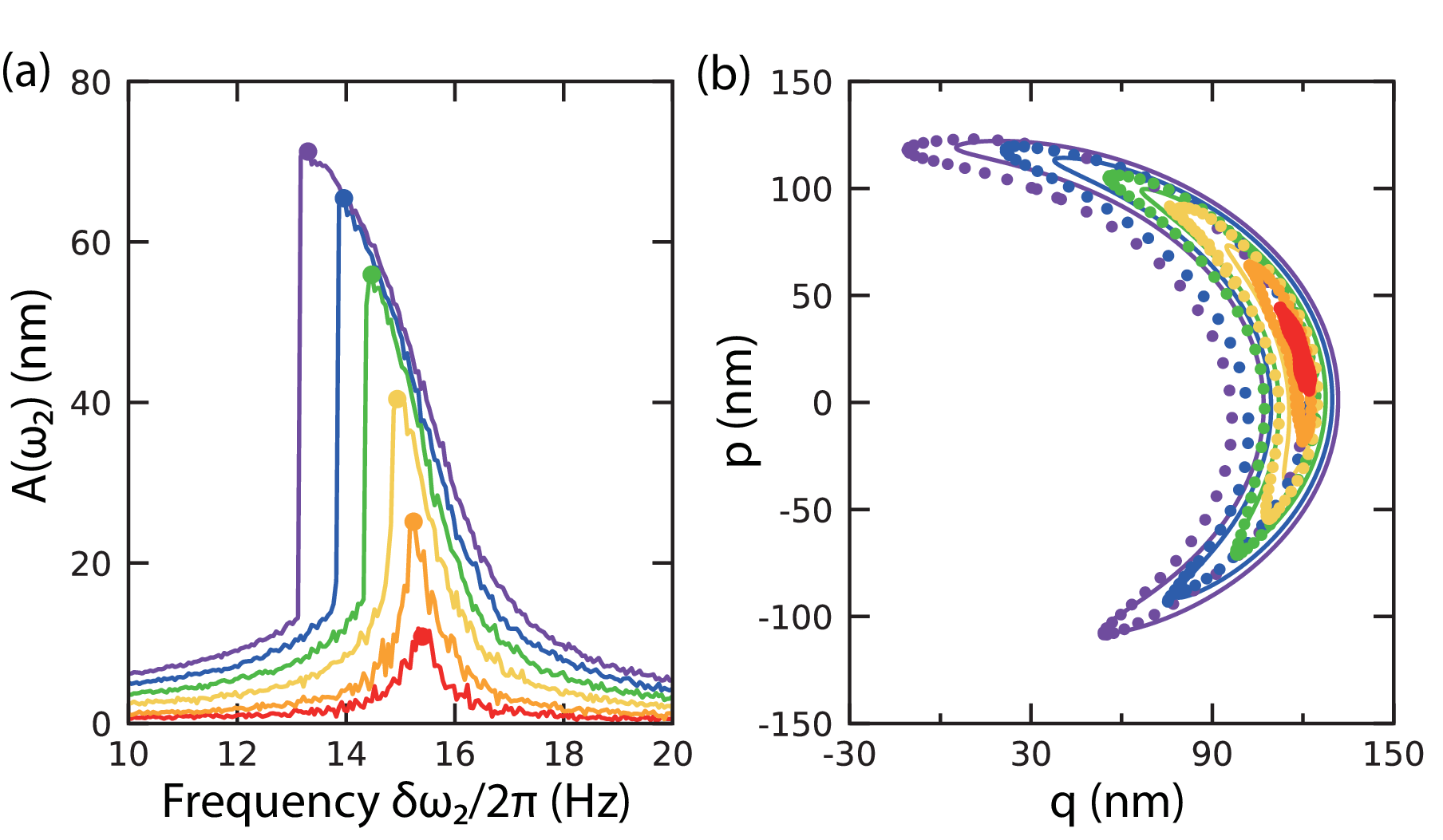}
\caption{(a) The amplitude $A(\omega_2)$ of vibrations  for downward frequency sweeps of the secondary-drive frequency $\omega_2$, measured at $F_2=15,~30,~60,~90,~120,$ and 150 $\rm m m/s^2$, with $F_1$ fixed at $\qty{3.0}{\meter / \second^2}$. The signal $A(\omega_2)$, is plotted against the detuning $\delta \omega_2=\omega_2 - \omega_1$ of the secondary drive. The peak values of the amplitudes for different $F_2$ are marked with solid circles. (b) The measured (circles) phase-space trajectories of ($q$, $p$) in the rotating frame for the frequencies $\omega_2$ corresponding to the maximal $A(\omega_2)$ in (a), with the matching color coding. The trajectories are obtained by homodyne measurement with reference frequency at $\omega_1$. Rainbow-colored curves are obtained by numerically solving the equations $\dot{q}=\partial_{p} \mathcal{H} -\Gamma q$ and $\dot{p}=-\partial_{q}\mathcal{H}-\Gamma p$ for
the corresponding sets of ($F_2$, $\delta \omega_2$).}
\label{fig:2}
\end{figure}

\begin{figure}[t]
\centering
\includegraphics[width=1\linewidth]{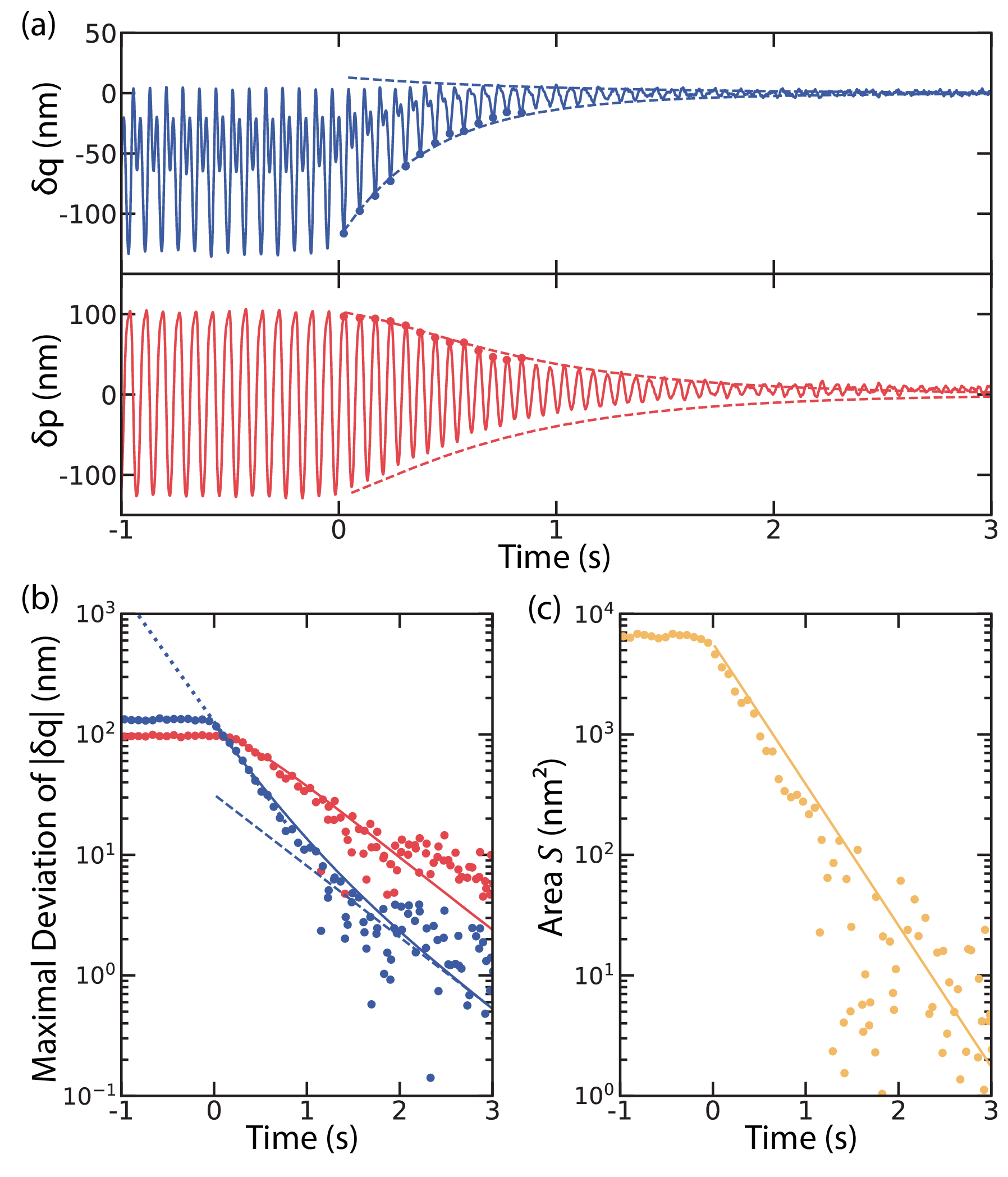}
\caption{(a) The ring-down signals of $\delta q=q(t)-q_{\text{eq}}$ and $\delta p =p(t)-p_{\text{eq}}$. The system is initially driven by the secondary drive with \( \delta \omega_2 /2\pi= \qty{13}{\hertz}\) and \( F_2 =\qty{150}{\milli \meter / \second^2} \), while the primary resonant drive is set at $\delta \omega_1/2\pi = \qty{86.9}{\hertz}$ and $F_1=\qty{3.0}{\meter / \second^2}$. \( F_2 \) is switched off at $t=0$. Dashed blue and red curves represent theoretical predictions for the ring-down envelopes of $\delta q$ and $\delta p$. (b) Log-linear plot of the maximal deviations for $\delta q$ (blue) and $\delta p$ (red). Solid curves show the theoretical predictions for the maximal displacement of the vibrations $|\delta q(t)|$ and $|\delta p(t)|$. The blue dotted and dashed lines represent the decay rate $2\Gamma $ and $\Gamma$, respectively. (c) The decay of the area enclosed by phase-space trajectories in the rotating frame over time. Yellow circles indicate the experimental data and the yellow curve is the  theoretical prediction.}
\label{fig:3}
\end{figure}

In Fig. 2 we show the measured response of the oscillator to the secondary drive. We focus on the response at the drive frequency $\omega_2$. For small $F_2$, the response is maximal at $\omega_2 =\omega_1 + \Omega$, and its amplitude $A(\omega_2)$ increases linearly with $F_2$. As the drive is further increased the response shows nonlinear behavior. Driving with $F_2$ beyond a critical value $F_{\rm cr_2}$ pushes the  response curve $A(\omega_2)$ into a bi-stable regime, and a set of downward frequency sweeping measurements reveals frequency pulling-away behavior, reminiscent of a softening Duffing response as shown in Fig.~\ref{fig:2}(a). This is caused by the dependence of the effective frequency of the vibrations in the rotating frame on the value of the Hamiltonian  $\mathcal{H}_0$. The small rainbow-colored circles in the figure indicate the maximal amplitudes at each value of $F_2$. They determine  the initial conditions of the ring-down measurements shown in Fig.~3, which are done by turning off $F_2$.

The phase-space trajectories of $(q,p)$ in the rotating frame are obtained by setting $\omega_2$ to produce the peak amplitude at each drive level of $F_2$ and then carrying out a homodyne detection with the the primary drive at $\omega_1$ as the reference; see Fig.~\ref{fig:2}(b). However, because of dissipation, the trajectories do not have mirror-reflection symmetry with respect to the axis $p=0$. This should be contrasted with the phase-space trajectories of self-sustained vibrations in the rotating frame for $F_2=0$ \cite{Ochs2022}. We did not observe such self-sustained vibrations.

\begin{figure}[t]
\centering
\includegraphics[width=1\linewidth]{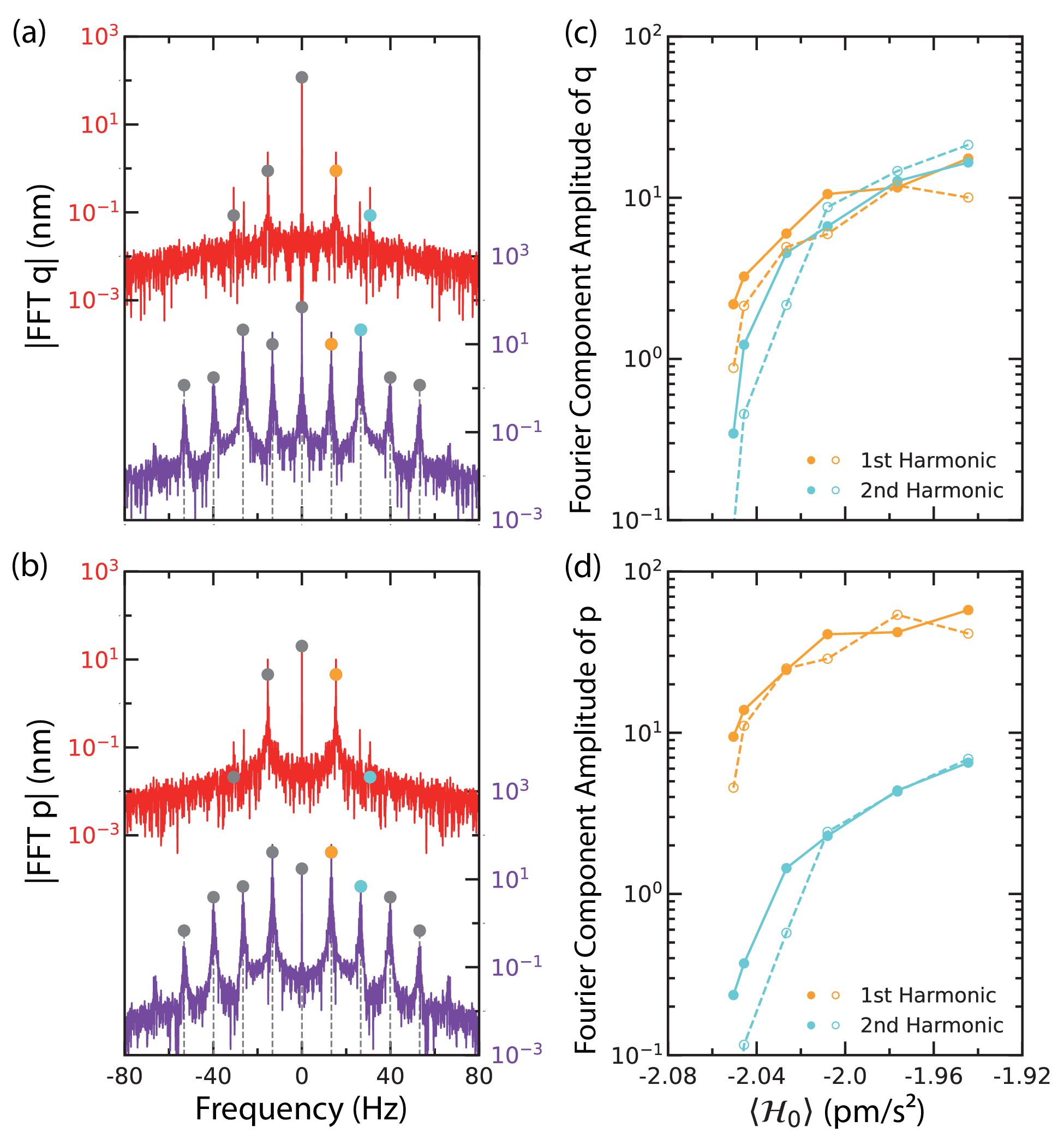}
\caption{(a),(b) Spectra of $q(t)$ and $p(t)$, obtained by fast Fourier tranform (FFT) of the outcomes of  homodyne measurements for two values of the secondary drive $(\delta\omega_2/2\pi, F_2)=(\qty{15.4}{\hertz}, \qty{15}{\milli \meter/\second^2})$ (upper red) and $(\delta\omega_2/2\pi, F_2)=(\qty{13.3}{\hertz}, \qty{150}{\milli \meter/\second^2})$ (lower purple). These values correspond to the two peak amplitudes marked by red and purple circles in Fig.~\ref{fig:2}(a), respectively, with $F_1$ also set to $\qty{3.0}{\meter / \second^2}$, the same values used for Fig.~\ref{fig:2} and Fig.~\ref{fig:3}. Colored circles (gray, yellow and cyan) at the peaks of the spectra are the theoretical values. (c),(d) Amplitudes of the first (yellow) and second (cyan) harmonics components of \(q(t) \) and \(p(t) \) as functions of the averaged Hamiltonian $\langle \mathcal{H}_0 \rangle$ calculated for  different $F_2$ and $\delta\omega_2/2\pi$. Filled circles and empty circles represent measured data and theoretical calculations, respectively. The solid and dashed lines between the circles are guides to the eye.}
\label{fig:4}
\end{figure}

Turning off the secondary drive \( F_2 \) induces ring-downs probed by homodyne measurements referenced at frequency $\omega_1$. As depicted in Fig.~\ref{fig:3}(a), the quadratures decay to the equilibrium position $ \delta q=q-q_{\text{eq}} \rightarrow 0 $, $ \delta p=p-p_{\text{eq}} \rightarrow 0 $. To characterize this decay process, we identify the points of the maximal deviations as the point furthest away from the equilibrium on the phase-space trajectories in Fig.~\ref{fig:2}(b) during each cycle. A handful of these points are marked by the blue and red circles on the curves in Fig.~\ref{fig:3}(a). The points of the maximal deviations are then plotted against time in Fig.~\ref{fig:3}(b) to illustrate the decay process. 

Two features immediately stand out from Fig.~\ref{fig:3}(b). The first is that the ring-down of $\delta q$ is not a simple exponential decay. Second, the decay processes of $q$ and $p$ are not identical, despite the fact that both quadratures in the dynamic model have the same damping coefficient $\Gamma$. The initial decay rates of $q$ and $p$ are measured to be \qty{2.50}{Hz} and \qty{1.31}{Hz}, respectively. These values are in a good agreement with the decay rates of $\approx 2\Gamma$ and $\approx \Gamma$ calculated for the chosen parameters and indicated by the early-time slopes in Fig. \ref{fig:3}(b). The decay rate of $q$ changes over time and eventually approaches $\Gamma$ in the small-amplitude limit (although the data in this limit is somewhat obscured by the noise), all the while the decay rate of $p$ remains constant. 

During the ringdown, because of the slow decay rate, the trajectories $q(t), p(t)$ in the rotating frame form a tight spiral. One can then consider the area $\mathcal{S}$ enclosed in a single turn of this spiral. In the absence of decay, this area would provide the action of conservative system with Hamiltonian $\mathcal{H}_0$. In the presence of decay, $\mathcal{S}$  decays as the mode approaches the fixed point. This decay is shown in Fig.~\ref{fig:3}(c). It is exponential at the rate of $\qty{2.69}{\hertz} \approx 2\Gamma $, consistent with our model. 
However, we emphasize, that the decay of the dynamical variables $q(t)$ and $p(t)$ in the rotating frame is not exponential. It is not a simple crossover typical of a double exponential decay either, previously seen in energy relaxations \cite{Güttinger2017}. Instead, the system exhibits a rather unusual non-exponential relaxation in which the two coupled quadratures relaxes differently from each other.

We now discuss the spectra of the system in the presence of a secondary drive. In Fig. \ref{fig:4} we show the Fourier transforms of the coordinate and momentum in the rotating frame. The spectra display multiple peaks, which correspond to the overtones of the vibrations at frequency $\delta\omega_2$, $q(t)=\sum q_{n}\exp(in\delta\omega_2t)$ and $p(t)=\sum p_{n}\exp(in\delta\omega_2t)$. The occurrence of the overtones leads to a frequency comb in the laboratory frame, with the spectral lines of the comb located at frequencies $\omega_1 + n\delta\omega_2$ with integer $n$. The comb is pronounced at a  sufficiently strong $F_2$. It  reveals a strong and nontrivial dependence of higher harmonics of $q(t)$ on the secondary drive. This dependence reflects the spectral composition of nonlinear  vibrations in the rotating frame as functions of the vibration ``energy'' $\mathcal{H}_0$ in the absence of decay and $F_2$. It is therefore instructive to describe it using the values  $\braket{\mathcal{H}_0}$, which are obtained by averaging $\mathcal{H}_0$ for given $F_2, \delta\omega_2$ along the  experimental trajectory.

It follows from the explicit form of the Hamiltonian trajectories in the absence of decay that, unexpectedly, already for comparatively small  $\mathcal{H}_0$, the second-order harmonic of $q(t)$ becomes comparable to the first-order harmonic, {\it i.e.} $q_2/q_1 \sim 1$, as shown in Fig.~\ref{fig:4}(c). In our device this happens for $\langle \mathcal{H}_0 \rangle \approx \qty{-1.98}{\pico m^2/s}$ or for the secondary drive amplitude $F_2 \approx \qty{120}{\milli \meter/\second^2}$, for $\delta\omega_2 = \qty{13.96}{\hertz}$. In contrast, for the same parameters, in the spectrum of $p(t)$ only the first harmonic is pronounced, see  Fig.~\ref{fig:4}(d). Such a difference in the behavior of two quadratures is distinct from the  response of most weakly nonlinear systems, such as a system with a standard Duffing nonlinearity, where the second harmonic remains small up to fairly large vibration  amplitudes, cf.\cite{Ochs2021resonant}. 

There is a connection between the evolution of the decay rates during the ring-down shown in Fig.~\ref{fig:3}(b) and the frequency content of the quadratures shown in Fig.~\ref{fig:4}. In a quasi-linear picture, the components  $q_n(t)$ and $p_n(t)$ decay at rates $n\Gamma$. Then, if the dominating components of $q(t)$ and $p(t)$ are $q_2$ and $p_1$, respectively, one might expect that $q(t)$ will decay at the rate $\sim 2\Gamma$ and $p(t)$ will decay at the rate $\sim \Gamma$. This argument provides an insight into the decay rates at large amplitudes, where $|q_2|\sim |q_1|$, whereas $|p_2|\ll |p_1|$. The non-exponential decay is only visible when the ring-down is initiated from relatively large $\langle \mathcal{H}_0 \rangle$, so that the decay rate of $q(t)$ changes from $\sim 2\Gamma$ to $\sim \Gamma$.

In conclusion, we experimentally investigated relaxation of a driven single-mode nonlinear nanomechanical resonator in the rotating frame. The nonlinearity is weak, and the vibration amplitude of the resonator decays exponentially in time, in the absence of a drive. However, we find that the decay toward the state of forced vibrations is unexpectedly complicated. The ring-down data that we collect shows  marked differences between the decays of the in-phase and quadrature components, which depart from a simple exponential law. Our results on the effect of a weak extra drive show  the onset of a pronounced second-harmonic signal in the rotating frame, but  only in the in-phase component. A minimal nonlinear model successfully describes the observations and reveals the relation between the non-exponential decay and the harmonic generation, which result from the interplay of the nonlinearity, dissipation, and the broken symmetry of the vibrations in the rotating frame.

More broadly, our findings reveal that even modest nonlinearities in the laboratory frame can become significantly amplified in the rotating frame, giving rise to qualitatively different dynamical behavior. This makes the rotating-frame method a sensitive and powerful tool for isolating and studying nonlinear effects—particularly in high-Q nanomechanical systems, where conventional linewidth measurements might mask important details. By connecting spectral signatures to time-domain decay patterns, our approach provides a framework for identifying and interpreting non-exponential relaxation processes, and controlling nonlinear phenomena. Beyond the present setting, this technique can also serve as a means of probing different dissipative environments
—for example, immersing the resonator in superfluid helium, where its vibration generates vortices\cite{Barquist2020, Guthrie2021, Barquist2022}—to study the nontrivial interplay between nonlinearity and dissipation mechanism, offering a pathway to deeper understanding.

\begin{acknowledgments}
This work is supported by National Research Foundation of Korea (NRF) grants (RS-2023-00207732 \& RS-2023-00256050). O.S. is grateful for support from the Israel Science Foundation (ISF) grant number 344/22 and the Pearlstone Center of Aeronautical Engineering Studies at BGU. J.S. acknowledges supports from NRF and IITP (RS-2023-00207732, RS-2022-00164799, RS-2024-00352688). SWS and MID acknowledge partial support from the Defense Advanced Research Projects Agency (DARPA) H6 program under cooperative agreement HR0011-23-2-0004.
\end{acknowledgments}

\bibliography{references}

\begin{thebibliography}{5}%
\makeatletter
\providecommand \@ifxundefined [1]{%
 \@ifx{#1\undefined}
}%
\providecommand \@ifnum [1]{%
 \ifnum #1\expandafter \@firstoftwo
 \else \expandafter \@secondoftwo
 \fi
}%
\providecommand \@ifx [1]{%
 \ifx #1\expandafter \@firstoftwo
 \else \expandafter \@secondoftwo
 \fi
}%
\providecommand \natexlab [1]{#1}%
\providecommand \enquote  [1]{``#1''}%
\providecommand \bibnamefont  [1]{#1}%
\providecommand \bibfnamefont [1]{#1}%
\providecommand \citenamefont [1]{#1}%
\providecommand \href@noop [0]{\@secondoftwo}%
\providecommand \href [0]{\begingroup \@sanitize@url \@href}%
\providecommand \@href[1]{\@@startlink{#1}\@@href}%
\providecommand \@@href[1]{\endgroup#1\@@endlink}%
\providecommand \@sanitize@url [0]{\catcode `\\12\catcode `\$12\catcode `\&12\catcode `\#12\catcode `\^12\catcode `\_12\catcode `\%12\relax}%
\providecommand \@@startlink[1]{}%
\providecommand \@@endlink[0]{}%
\providecommand \url  [0]{\begingroup\@sanitize@url \@url }%
\providecommand \@url [1]{\endgroup\@href {#1}{\urlprefix }}%
\providecommand \urlprefix  [0]{URL }%
\providecommand \Eprint [0]{\href }%
\providecommand \doibase [0]{https://doi.org/}%
\providecommand \selectlanguage [0]{\@gobble}%
\providecommand \bibinfo  [0]{\@secondoftwo}%
\providecommand \bibfield  [0]{\@secondoftwo}%
\providecommand \translation [1]{[#1]}%
\providecommand \BibitemOpen [0]{}%
\providecommand \bibitemStop [0]{}%
\providecommand \bibitemNoStop [0]{.\EOS\space}%
\providecommand \EOS [0]{\spacefactor3000\relax}%
\providecommand \BibitemShut  [1]{\csname bibitem#1\endcsname}%
\let\auto@bib@innerbib\@empty
\bibitem [{\citenamefont {Guckenheimer}\ and\ \citenamefont {Holmes}(2013)}]{Guckenheimer2013}%
  \BibitemOpen
  \bibfield  {author} {\bibinfo {author} {\bibfnamefont {J.}~\bibnamefont {Guckenheimer}}\ and\ \bibinfo {author} {\bibfnamefont {P.~J.}\ \bibnamefont {Holmes}},\ }\href@noop {} {\emph {\bibinfo {title} {Nonlinear Oscillations, Dynamical Systems, and Bifurcations of Vector Fields}}}\ (\bibinfo  {publisher} {Springer},\ \bibinfo {address} {New York},\ \bibinfo {year} {2013})\BibitemShut {NoStop}%
\bibitem [{\citenamefont {Cleland}\ and\ \citenamefont {Roukes}(1996)}]{Cleland1996}%
  \BibitemOpen
  \bibfield  {author} {\bibinfo {author} {\bibfnamefont {A.~N.}\ \bibnamefont {Cleland}}\ and\ \bibinfo {author} {\bibfnamefont {M.~L.}\ \bibnamefont {Roukes}},\ }\bibfield  {title} {\bibinfo {title} {Fabrication of high frequency nanometer scale mechanical resonators from bulk si crystals},\ }\href {https://doi.org/10.1063/1.117723} {\bibfield  {journal} {\bibinfo  {journal} {Applied Physics Letters}\ }\textbf {\bibinfo {volume} {69}},\ \bibinfo {pages} {2653} (\bibinfo {year} {1996})}\BibitemShut {NoStop}%
\bibitem [{\citenamefont {Maillet}\ \emph {et~al.}(2016)\citenamefont {Maillet}, \citenamefont {Vavrek}, \citenamefont {Fefferman}, \citenamefont {Bourgeois},\ and\ \citenamefont {Collin}}]{maillet2016}%
  \BibitemOpen
  \bibfield  {author} {\bibinfo {author} {\bibfnamefont {O.}~\bibnamefont {Maillet}}, \bibinfo {author} {\bibfnamefont {F.}~\bibnamefont {Vavrek}}, \bibinfo {author} {\bibfnamefont {A.~D.}\ \bibnamefont {Fefferman}}, \bibinfo {author} {\bibfnamefont {O.}~\bibnamefont {Bourgeois}},\ and\ \bibinfo {author} {\bibfnamefont {E.}~\bibnamefont {Collin}},\ }\bibfield  {title} {\bibinfo {title} {{Classical decoherence in a nanomechanical resonator}},\ }\href {https://doi.org/10.1088/1367-2630/18/7/073022} {\bibfield  {journal} {\bibinfo  {journal} {New Journal of Physics}\ }\textbf {\bibinfo {volume} {18}},\ \bibinfo {pages} {073022} (\bibinfo {year} {2016})},\ \Eprint {https://arxiv.org/abs/1511.02120} {1511.02120} \BibitemShut {NoStop}%
\bibitem [{\citenamefont {{Zurich Instruments}}(2023)}]{Zurich_lock_in}%
  \BibitemOpen
  \bibfield  {author} {\bibinfo {author} {\bibnamefont {{Zurich Instruments}}},\ }\href@noop {} {\emph {\bibinfo {title} {Principles of Lock-in Detection}}},\ \bibinfo {type} {Tech. Rep.}\ (\bibinfo  {institution} {Zurich Instruments AG},\ \bibinfo {address} {Zurich, Switzerland},\ \bibinfo {year} {2023})\BibitemShut {NoStop}%
\bibitem [{\citenamefont {Kumar}\ \emph {et~al.}(2021)\citenamefont {Kumar}, \citenamefont {Rebari}, \citenamefont {Pal}, \citenamefont {Yadav}, \citenamefont {Kumar}, \citenamefont {Aggarwal}, \citenamefont {Indrajeet},\ and\ \citenamefont {Venkatesan}}]{nano_tune}%
  \BibitemOpen
  \bibfield  {author} {\bibinfo {author} {\bibfnamefont {S.}~\bibnamefont {Kumar}}, \bibinfo {author} {\bibfnamefont {S.}~\bibnamefont {Rebari}}, \bibinfo {author} {\bibfnamefont {S.~P.}\ \bibnamefont {Pal}}, \bibinfo {author} {\bibfnamefont {S.~S.}\ \bibnamefont {Yadav}}, \bibinfo {author} {\bibfnamefont {A.}~\bibnamefont {Kumar}}, \bibinfo {author} {\bibfnamefont {A.}~\bibnamefont {Aggarwal}}, \bibinfo {author} {\bibfnamefont {S.}~\bibnamefont {Indrajeet}},\ and\ \bibinfo {author} {\bibfnamefont {A.}~\bibnamefont {Venkatesan}},\ }\bibfield  {title} {\bibinfo {title} {{Temperature-Dependent Nonlinear Damping in Palladium Nanomechanical Resonators}},\ }\href {https://doi.org/10.1021/acs.nanolett.1c00109} {\bibfield  {journal} {\bibinfo  {journal} {Nano Letters}\ }\textbf {\bibinfo {volume} {21}},\ \bibinfo {pages} {2975} (\bibinfo {year} {2021})},\ \Eprint {https://arxiv.org/abs/2009.08324} {2009.08324} \BibitemShut {NoStop}%
\end{thebibliography}%


\begin{thebibliography}{43}%
\makeatletter
\providecommand \@ifxundefined [1]{%
 \@ifx{#1\undefined}
}%
\providecommand \@ifnum [1]{%
 \ifnum #1\expandafter \@firstoftwo
 \else \expandafter \@secondoftwo
 \fi
}%
\providecommand \@ifx [1]{%
 \ifx #1\expandafter \@firstoftwo
 \else \expandafter \@secondoftwo
 \fi
}%
\providecommand \natexlab [1]{#1}%
\providecommand \enquote  [1]{``#1''}%
\providecommand \bibnamefont  [1]{#1}%
\providecommand \bibfnamefont [1]{#1}%
\providecommand \citenamefont [1]{#1}%
\providecommand \href@noop [0]{\@secondoftwo}%
\providecommand \href [0]{\begingroup \@sanitize@url \@href}%
\providecommand \@href[1]{\@@startlink{#1}\@@href}%
\providecommand \@@href[1]{\endgroup#1\@@endlink}%
\providecommand \@sanitize@url [0]{\catcode `\\12\catcode `\$12\catcode `\&12\catcode `\#12\catcode `\^12\catcode `\_12\catcode `\%12\relax}%
\providecommand \@@startlink[1]{}%
\providecommand \@@endlink[0]{}%
\providecommand \url  [0]{\begingroup\@sanitize@url \@url }%
\providecommand \@url [1]{\endgroup\@href {#1}{\urlprefix }}%
\providecommand \urlprefix  [0]{URL }%
\providecommand \Eprint [0]{\href }%
\providecommand \doibase [0]{https://doi.org/}%
\providecommand \selectlanguage [0]{\@gobble}%
\providecommand \bibinfo  [0]{\@secondoftwo}%
\providecommand \bibfield  [0]{\@secondoftwo}%
\providecommand \translation [1]{[#1]}%
\providecommand \BibitemOpen [0]{}%
\providecommand \bibitemStop [0]{}%
\providecommand \bibitemNoStop [0]{.\EOS\space}%
\providecommand \EOS [0]{\spacefactor3000\relax}%
\providecommand \BibitemShut  [1]{\csname bibitem#1\endcsname}%
\let\auto@bib@innerbib\@empty
\bibitem [{\citenamefont {Ekinci}\ and\ \citenamefont {Roukes}(2005)}]{Ekinci2005}%
  \BibitemOpen
  \bibfield  {author} {\bibinfo {author} {\bibfnamefont {K.~L.}\ \bibnamefont {Ekinci}}\ and\ \bibinfo {author} {\bibfnamefont {M.~L.}\ \bibnamefont {Roukes}},\ }\bibfield  {title} {\bibinfo {title} {Nanoelectromechanical systems},\ }\href {https://doi.org/10.1063/1.1927327} {\bibfield  {journal} {\bibinfo  {journal} {Review of Scientific Instruments}\ }\textbf {\bibinfo {volume} {76}},\ \bibinfo {pages} {061101} (\bibinfo {year} {2005})}\BibitemShut {NoStop}%
\bibitem [{\citenamefont {Unterreithmeier}\ \emph {et~al.}(2010)\citenamefont {Unterreithmeier}, \citenamefont {Faust},\ and\ \citenamefont {Kotthaus}}]{Unterreithmeier2010}%
  \BibitemOpen
  \bibfield  {author} {\bibinfo {author} {\bibfnamefont {Q.~P.}\ \bibnamefont {Unterreithmeier}}, \bibinfo {author} {\bibfnamefont {T.}~\bibnamefont {Faust}},\ and\ \bibinfo {author} {\bibfnamefont {J.~P.}\ \bibnamefont {Kotthaus}},\ }\bibfield  {title} {\bibinfo {title} {{Damping of Nanomechanical Resonators}},\ }\href {https://doi.org/10.1103/physrevlett.105.027205} {\bibfield  {journal} {\bibinfo  {journal} {Physical Review Letters}\ }\textbf {\bibinfo {volume} {105}},\ \bibinfo {pages} {027205} (\bibinfo {year} {2010})},\ \Eprint {https://arxiv.org/abs/1003.1868} {1003.1868} \BibitemShut {NoStop}%
\bibitem [{\citenamefont {Chen}\ \emph {et~al.}(2017)\citenamefont {Chen}, \citenamefont {Zanette}, \citenamefont {Czaplewski}, \citenamefont {Shaw},\ and\ \citenamefont {López}}]{chen2017}%
  \BibitemOpen
  \bibfield  {author} {\bibinfo {author} {\bibfnamefont {C.}~\bibnamefont {Chen}}, \bibinfo {author} {\bibfnamefont {D.~H.}\ \bibnamefont {Zanette}}, \bibinfo {author} {\bibfnamefont {D.~A.}\ \bibnamefont {Czaplewski}}, \bibinfo {author} {\bibfnamefont {S.}~\bibnamefont {Shaw}},\ and\ \bibinfo {author} {\bibfnamefont {D.}~\bibnamefont {López}},\ }\bibfield  {title} {\bibinfo {title} {{Direct observation of coherent energy transfer in nonlinear micromechanical oscillators}},\ }\href {https://doi.org/10.1038/ncomms15523} {\bibfield  {journal} {\bibinfo  {journal} {Nature Communications}\ }\textbf {\bibinfo {volume} {8}},\ \bibinfo {pages} {15523} (\bibinfo {year} {2017})},\ \Eprint {https://arxiv.org/abs/1612.00490} {1612.00490} \BibitemShut {NoStop}%
\bibitem [{\citenamefont {Clerk}\ \emph {et~al.}(2010)\citenamefont {Clerk}, \citenamefont {Devoret}, \citenamefont {Girvin}, \citenamefont {Marquardt},\ and\ \citenamefont {Schoelkopf}}]{Clerk2010}%
  \BibitemOpen
  \bibfield  {author} {\bibinfo {author} {\bibfnamefont {A.~A.}\ \bibnamefont {Clerk}}, \bibinfo {author} {\bibfnamefont {M.~H.}\ \bibnamefont {Devoret}}, \bibinfo {author} {\bibfnamefont {S.~M.}\ \bibnamefont {Girvin}}, \bibinfo {author} {\bibfnamefont {F.}~\bibnamefont {Marquardt}},\ and\ \bibinfo {author} {\bibfnamefont {R.~J.}\ \bibnamefont {Schoelkopf}},\ }\bibfield  {title} {\bibinfo {title} {Introduction to quantum noise, measurement, and amplification},\ }\href {https://doi.org/10.1103/RevModPhys.82.1155} {\bibfield  {journal} {\bibinfo  {journal} {Rev. Mod. Phys.}\ }\textbf {\bibinfo {volume} {82}},\ \bibinfo {pages} {1155} (\bibinfo {year} {2010})}\BibitemShut {NoStop}%
\bibitem [{\citenamefont {Bachtold}\ \emph {et~al.}(2022)\citenamefont {Bachtold}, \citenamefont {Moser},\ and\ \citenamefont {Dykman}}]{bachtold2022}%
  \BibitemOpen
  \bibfield  {author} {\bibinfo {author} {\bibfnamefont {A.}~\bibnamefont {Bachtold}}, \bibinfo {author} {\bibfnamefont {J.}~\bibnamefont {Moser}},\ and\ \bibinfo {author} {\bibfnamefont {M.~I.}\ \bibnamefont {Dykman}},\ }\bibfield  {title} {\bibinfo {title} {{Mesoscopic physics of nanomechanical systems}},\ }\href {https://doi.org/10.1103/revmodphys.94.045005} {\bibfield  {journal} {\bibinfo  {journal} {Reviews of Modern Physics}\ }\textbf {\bibinfo {volume} {94}},\ \bibinfo {pages} {045005} (\bibinfo {year} {2022})}\BibitemShut {NoStop}%
\bibitem [{\citenamefont {Maillet}\ \emph {et~al.}(2017)\citenamefont {Maillet}, \citenamefont {Zhou}, \citenamefont {Gazizulin}, \citenamefont {Cid}, \citenamefont {Defoort}, \citenamefont {Bourgeois},\ and\ \citenamefont {Collin}}]{Maillet2017}%
  \BibitemOpen
  \bibfield  {author} {\bibinfo {author} {\bibfnamefont {O.}~\bibnamefont {Maillet}}, \bibinfo {author} {\bibfnamefont {X.}~\bibnamefont {Zhou}}, \bibinfo {author} {\bibfnamefont {R.}~\bibnamefont {Gazizulin}}, \bibinfo {author} {\bibfnamefont {A.~M.}\ \bibnamefont {Cid}}, \bibinfo {author} {\bibfnamefont {M.}~\bibnamefont {Defoort}}, \bibinfo {author} {\bibfnamefont {O.}~\bibnamefont {Bourgeois}},\ and\ \bibinfo {author} {\bibfnamefont {E.}~\bibnamefont {Collin}},\ }\bibfield  {title} {\bibinfo {title} {Nonlinear frequency transduction of nanomechanical brownian motion},\ }\href {https://doi.org/10.1103/PhysRevB.96.165434} {\bibfield  {journal} {\bibinfo  {journal} {Phys. Rev. B}\ }\textbf {\bibinfo {volume} {96}},\ \bibinfo {pages} {165434} (\bibinfo {year} {2017})}\BibitemShut {NoStop}%
\bibitem [{\citenamefont {Zhang}\ and\ \citenamefont {Dykman}(2015)}]{Zhang2015}%
  \BibitemOpen
  \bibfield  {author} {\bibinfo {author} {\bibfnamefont {Y.}~\bibnamefont {Zhang}}\ and\ \bibinfo {author} {\bibfnamefont {M.~I.}\ \bibnamefont {Dykman}},\ }\bibfield  {title} {\bibinfo {title} {Spectral effects of dispersive mode coupling in driven mesoscopic systems},\ }\href {https://doi.org/10.1103/PhysRevB.92.165419} {\bibfield  {journal} {\bibinfo  {journal} {Phys. Rev. B}\ }\textbf {\bibinfo {volume} {92}},\ \bibinfo {pages} {165419} (\bibinfo {year} {2015})}\BibitemShut {NoStop}%
\bibitem [{\citenamefont {Fong}\ \emph {et~al.}(2012)\citenamefont {Fong}, \citenamefont {Pernice},\ and\ \citenamefont {Tang}}]{Fong2012}%
  \BibitemOpen
  \bibfield  {author} {\bibinfo {author} {\bibfnamefont {K.~Y.}\ \bibnamefont {Fong}}, \bibinfo {author} {\bibfnamefont {W.~H.~P.}\ \bibnamefont {Pernice}},\ and\ \bibinfo {author} {\bibfnamefont {H.~X.}\ \bibnamefont {Tang}},\ }\bibfield  {title} {\bibinfo {title} {Frequency and phase noise of ultrahigh $q$ silicon nitride nanomechanical resonators},\ }\href {https://doi.org/10.1103/PhysRevB.85.161410} {\bibfield  {journal} {\bibinfo  {journal} {Phys. Rev. B}\ }\textbf {\bibinfo {volume} {85}},\ \bibinfo {pages} {161410} (\bibinfo {year} {2012})}\BibitemShut {NoStop}%
\bibitem [{\citenamefont {Huang}\ \emph {et~al.}(2019)\citenamefont {Huang}, \citenamefont {Soskin}, \citenamefont {Khovanov}, \citenamefont {Mannella}, \citenamefont {Ninios},\ and\ \citenamefont {Chan}}]{Huang2019}%
  \BibitemOpen
  \bibfield  {author} {\bibinfo {author} {\bibfnamefont {L.}~\bibnamefont {Huang}}, \bibinfo {author} {\bibfnamefont {S.~M.}\ \bibnamefont {Soskin}}, \bibinfo {author} {\bibfnamefont {I.~A.}\ \bibnamefont {Khovanov}}, \bibinfo {author} {\bibfnamefont {R.}~\bibnamefont {Mannella}}, \bibinfo {author} {\bibfnamefont {K.}~\bibnamefont {Ninios}},\ and\ \bibinfo {author} {\bibfnamefont {H.~B.}\ \bibnamefont {Chan}},\ }\bibfield  {title} {\bibinfo {title} {{Frequency stabilization and noise-induced spectral narrowing in resonators with zero dispersion}},\ }\href {https://doi.org/10.1038/s41467-019-11946-8} {\bibfield  {journal} {\bibinfo  {journal} {Nature Communications}\ }\textbf {\bibinfo {volume} {10}},\ \bibinfo {pages} {3930} (\bibinfo {year} {2019})},\ \Eprint {https://arxiv.org/abs/1909.01090} {1909.01090} \BibitemShut {NoStop}%
\bibitem [{\citenamefont {Schmid}\ \emph {et~al.}(2011)\citenamefont {Schmid}, \citenamefont {Jensen}, \citenamefont {Nielsen},\ and\ \citenamefont {Boisen}}]{schmid2011}%
  \BibitemOpen
  \bibfield  {author} {\bibinfo {author} {\bibfnamefont {S.}~\bibnamefont {Schmid}}, \bibinfo {author} {\bibfnamefont {K.~D.}\ \bibnamefont {Jensen}}, \bibinfo {author} {\bibfnamefont {K.~H.}\ \bibnamefont {Nielsen}},\ and\ \bibinfo {author} {\bibfnamefont {A.}~\bibnamefont {Boisen}},\ }\bibfield  {title} {\bibinfo {title} {Damping mechanisms in high-$q$ micro and nanomechanical string resonators},\ }\href {https://doi.org/10.1103/PhysRevB.84.165307} {\bibfield  {journal} {\bibinfo  {journal} {Phys. Rev. B}\ }\textbf {\bibinfo {volume} {84}},\ \bibinfo {pages} {165307} (\bibinfo {year} {2011})}\BibitemShut {NoStop}%
\bibitem [{\citenamefont {Güttinger}\ \emph {et~al.}(2017)\citenamefont {Güttinger}, \citenamefont {Noury}, \citenamefont {Weber}, \citenamefont {Eriksson}, \citenamefont {Lagoin}, \citenamefont {Moser}, \citenamefont {Eichler}, \citenamefont {Wallraff}, \citenamefont {Isacsson},\ and\ \citenamefont {Bachtold}}]{Güttinger2017}%
  \BibitemOpen
  \bibfield  {author} {\bibinfo {author} {\bibfnamefont {J.}~\bibnamefont {Güttinger}}, \bibinfo {author} {\bibfnamefont {A.}~\bibnamefont {Noury}}, \bibinfo {author} {\bibfnamefont {P.}~\bibnamefont {Weber}}, \bibinfo {author} {\bibfnamefont {A.~M.}\ \bibnamefont {Eriksson}}, \bibinfo {author} {\bibfnamefont {C.}~\bibnamefont {Lagoin}}, \bibinfo {author} {\bibfnamefont {J.}~\bibnamefont {Moser}}, \bibinfo {author} {\bibfnamefont {C.}~\bibnamefont {Eichler}}, \bibinfo {author} {\bibfnamefont {A.}~\bibnamefont {Wallraff}}, \bibinfo {author} {\bibfnamefont {A.}~\bibnamefont {Isacsson}},\ and\ \bibinfo {author} {\bibfnamefont {A.}~\bibnamefont {Bachtold}},\ }\bibfield  {title} {\bibinfo {title} {{Energy-dependent path of dissipation in nanomechanical resonators}},\ }\href {https://doi.org/10.1038/nnano.2017.86} {\bibfield  {journal} {\bibinfo  {journal} {Nature Nanotechnology}\ }\textbf {\bibinfo {volume} {12}},\ \bibinfo {pages} {631} (\bibinfo {year} {2017})}\BibitemShut {NoStop}%
\bibitem [{\citenamefont {Wang}\ \emph {et~al.}(2022{\natexlab{a}})\citenamefont {Wang}, \citenamefont {Perez-Morelo}, \citenamefont {Lopez},\ and\ \citenamefont {Aksyuk}}]{Wang2022}%
  \BibitemOpen
  \bibfield  {author} {\bibinfo {author} {\bibfnamefont {M.}~\bibnamefont {Wang}}, \bibinfo {author} {\bibfnamefont {D.~J.}\ \bibnamefont {Perez-Morelo}}, \bibinfo {author} {\bibfnamefont {D.}~\bibnamefont {Lopez}},\ and\ \bibinfo {author} {\bibfnamefont {V.~A.}\ \bibnamefont {Aksyuk}},\ }\bibfield  {title} {\bibinfo {title} {{Persistent Nonlinear Phase-Locking and Nonmonotonic Energy Dissipation in Micromechanical Resonators}},\ }\href {https://doi.org/10.1103/physrevx.12.041025} {\bibfield  {journal} {\bibinfo  {journal} {Physical Review X}\ }\textbf {\bibinfo {volume} {12}},\ \bibinfo {pages} {041025} (\bibinfo {year} {2022}{\natexlab{a}})},\ \Eprint {https://arxiv.org/abs/2206.01089} {2206.01089} \BibitemShut {NoStop}%
\bibitem [{\citenamefont {Shoshani}\ \emph {et~al.}(2017)\citenamefont {Shoshani}, \citenamefont {Shaw},\ and\ \citenamefont {Dykman}}]{Shoshani2017}%
  \BibitemOpen
  \bibfield  {author} {\bibinfo {author} {\bibfnamefont {O.}~\bibnamefont {Shoshani}}, \bibinfo {author} {\bibfnamefont {S.~W.}\ \bibnamefont {Shaw}},\ and\ \bibinfo {author} {\bibfnamefont {M.~I.}\ \bibnamefont {Dykman}},\ }\bibfield  {title} {\bibinfo {title} {{Anomalous Decay of Nanomechanical Modes Going Through Nonlinear Resonance}},\ }\href {https://doi.org/10.1038/s41598-017-17184-6} {\bibfield  {journal} {\bibinfo  {journal} {Scientific Reports}\ }\textbf {\bibinfo {volume} {7}},\ \bibinfo {pages} {18091} (\bibinfo {year} {2017})}\BibitemShut {NoStop}%
\bibitem [{\citenamefont {Schneider}\ \emph {et~al.}(2014)\citenamefont {Schneider}, \citenamefont {Singh}, \citenamefont {Venstra}, \citenamefont {Meerwaldt},\ and\ \citenamefont {Steele}}]{Schneider2014}%
  \BibitemOpen
  \bibfield  {author} {\bibinfo {author} {\bibfnamefont {B.~H.}\ \bibnamefont {Schneider}}, \bibinfo {author} {\bibfnamefont {V.}~\bibnamefont {Singh}}, \bibinfo {author} {\bibfnamefont {W.~J.}\ \bibnamefont {Venstra}}, \bibinfo {author} {\bibfnamefont {H.~B.}\ \bibnamefont {Meerwaldt}},\ and\ \bibinfo {author} {\bibfnamefont {G.~A.}\ \bibnamefont {Steele}},\ }\bibfield  {title} {\bibinfo {title} {{Observation of decoherence in a carbon nanotube mechanical resonator}},\ }\href {https://doi.org/10.1038/ncomms6819} {\bibfield  {journal} {\bibinfo  {journal} {Nature Communications}\ }\textbf {\bibinfo {volume} {5}},\ \bibinfo {pages} {5819} (\bibinfo {year} {2014})},\ \Eprint {https://arxiv.org/abs/1503.06815} {1503.06815} \BibitemShut {NoStop}%
\bibitem [{\citenamefont {Eichler}\ \emph {et~al.}(2011)\citenamefont {Eichler}, \citenamefont {Moser}, \citenamefont {Chaste}, \citenamefont {Zdrojek}, \citenamefont {Wilson-Rae},\ and\ \citenamefont {Bachtold}}]{Eichler2011}%
  \BibitemOpen
  \bibfield  {author} {\bibinfo {author} {\bibfnamefont {A.}~\bibnamefont {Eichler}}, \bibinfo {author} {\bibfnamefont {J.}~\bibnamefont {Moser}}, \bibinfo {author} {\bibfnamefont {J.}~\bibnamefont {Chaste}}, \bibinfo {author} {\bibfnamefont {M.}~\bibnamefont {Zdrojek}}, \bibinfo {author} {\bibfnamefont {I.}~\bibnamefont {Wilson-Rae}},\ and\ \bibinfo {author} {\bibfnamefont {A.}~\bibnamefont {Bachtold}},\ }\bibfield  {title} {\bibinfo {title} {{Nonlinear damping in mechanical resonators made from carbon nanotubes and graphene}},\ }\href {https://doi.org/10.1038/nnano.2011.71} {\bibfield  {journal} {\bibinfo  {journal} {Nature Nanotechnology}\ }\textbf {\bibinfo {volume} {6}},\ \bibinfo {pages} {339} (\bibinfo {year} {2011})},\ \Eprint {https://arxiv.org/abs/1103.1788} {1103.1788} \BibitemShut {NoStop}%
\bibitem [{\citenamefont {Polunin}\ \emph {et~al.}(2016)\citenamefont {Polunin}, \citenamefont {Yang}, \citenamefont {Dykman}, \citenamefont {Kenny},\ and\ \citenamefont {Shaw}}]{Polunin2016}%
  \BibitemOpen
  \bibfield  {author} {\bibinfo {author} {\bibfnamefont {P.~M.}\ \bibnamefont {Polunin}}, \bibinfo {author} {\bibfnamefont {Y.}~\bibnamefont {Yang}}, \bibinfo {author} {\bibfnamefont {M.~I.}\ \bibnamefont {Dykman}}, \bibinfo {author} {\bibfnamefont {T.~W.}\ \bibnamefont {Kenny}},\ and\ \bibinfo {author} {\bibfnamefont {S.~W.}\ \bibnamefont {Shaw}},\ }\bibfield  {title} {\bibinfo {title} {Characterization of mems resonator nonlinearities using the ringdown response},\ }\href {https://doi.org/10.1109/JMEMS.2016.2529296} {\bibfield  {journal} {\bibinfo  {journal} {Journal of Microelectromechanical Systems}\ }\textbf {\bibinfo {volume} {25}},\ \bibinfo {pages} {297} (\bibinfo {year} {2016})}\BibitemShut {NoStop}%
\bibitem [{\citenamefont {Huber}\ \emph {et~al.}(2020)\citenamefont {Huber}, \citenamefont {Rastelli}, \citenamefont {Seitner}, \citenamefont {Kölbl}, \citenamefont {Belzig}, \citenamefont {Dykman},\ and\ \citenamefont {Weig}}]{Huber2020}%
  \BibitemOpen
  \bibfield  {author} {\bibinfo {author} {\bibfnamefont {J.~S.}\ \bibnamefont {Huber}}, \bibinfo {author} {\bibfnamefont {G.}~\bibnamefont {Rastelli}}, \bibinfo {author} {\bibfnamefont {M.~J.}\ \bibnamefont {Seitner}}, \bibinfo {author} {\bibfnamefont {J.}~\bibnamefont {Kölbl}}, \bibinfo {author} {\bibfnamefont {W.}~\bibnamefont {Belzig}}, \bibinfo {author} {\bibfnamefont {M.~I.}\ \bibnamefont {Dykman}},\ and\ \bibinfo {author} {\bibfnamefont {E.~M.}\ \bibnamefont {Weig}},\ }\bibfield  {title} {\bibinfo {title} {{Spectral Evidence of Squeezing of a Weakly Damped Driven Nanomechanical Mode}},\ }\href {https://doi.org/10.1103/physrevx.10.021066} {\bibfield  {journal} {\bibinfo  {journal} {Physical Review X}\ }\textbf {\bibinfo {volume} {10}},\ \bibinfo {pages} {021066} (\bibinfo {year} {2020})},\ \Eprint {https://arxiv.org/abs/1903.07601} {1903.07601} \BibitemShut {NoStop}%
\bibitem [{\citenamefont {Ochs}\ \emph {et~al.}(2021{\natexlab{a}})\citenamefont {Ochs}, \citenamefont {Seitner}, \citenamefont {Dykman},\ and\ \citenamefont {Weig}}]{Ochs2021}%
  \BibitemOpen
  \bibfield  {author} {\bibinfo {author} {\bibfnamefont {J.~S.}\ \bibnamefont {Ochs}}, \bibinfo {author} {\bibfnamefont {M.}~\bibnamefont {Seitner}}, \bibinfo {author} {\bibfnamefont {M.~I.}\ \bibnamefont {Dykman}},\ and\ \bibinfo {author} {\bibfnamefont {E.~M.}\ \bibnamefont {Weig}},\ }\bibfield  {title} {\bibinfo {title} {Amplification and spectral evidence of squeezing in the response of a strongly driven nanoresonator to a probe field},\ }\href {https://doi.org/10.1103/physreva.103.013506} {\bibfield  {journal} {\bibinfo  {journal} {Physical Review A}\ }\textbf {\bibinfo {volume} {103}},\ \bibinfo {pages} {013506} (\bibinfo {year} {2021}{\natexlab{a}})},\ \Eprint {https://arxiv.org/abs/2007.15382} {2007.15382} \BibitemShut {NoStop}%
\bibitem [{\citenamefont {Yang}\ \emph {et~al.}(2021)\citenamefont {Yang}, \citenamefont {Fu}, \citenamefont {Bosnjak}, \citenamefont {Blick}, \citenamefont {Jiang},\ and\ \citenamefont {Scheer}}]{Yang2021}%
  \BibitemOpen
  \bibfield  {author} {\bibinfo {author} {\bibfnamefont {F.}~\bibnamefont {Yang}}, \bibinfo {author} {\bibfnamefont {M.}~\bibnamefont {Fu}}, \bibinfo {author} {\bibfnamefont {B.}~\bibnamefont {Bosnjak}}, \bibinfo {author} {\bibfnamefont {R.~H.}\ \bibnamefont {Blick}}, \bibinfo {author} {\bibfnamefont {Y.}~\bibnamefont {Jiang}},\ and\ \bibinfo {author} {\bibfnamefont {E.}~\bibnamefont {Scheer}},\ }\bibfield  {title} {\bibinfo {title} {{Mechanically Modulated Sideband and Squeezing Effects of Membrane Resonators}},\ }\href {https://doi.org/10.1103/physrevlett.127.184301} {\bibfield  {journal} {\bibinfo  {journal} {Physical Review Letters}\ }\textbf {\bibinfo {volume} {127}},\ \bibinfo {pages} {184301} (\bibinfo {year} {2021})},\ \Eprint {https://arxiv.org/abs/2107.10355} {2107.10355} \BibitemShut {NoStop}%
\bibitem [{\citenamefont {Zhang}\ \emph {et~al.}(2024)\citenamefont {Zhang}, \citenamefont {Yan}, \citenamefont {Dong}, \citenamefont {Dykman},\ and\ \citenamefont {Chan}}]{Zhang2024}%
  \BibitemOpen
  \bibfield  {author} {\bibinfo {author} {\bibfnamefont {B.}~\bibnamefont {Zhang}}, \bibinfo {author} {\bibfnamefont {Y.}~\bibnamefont {Yan}}, \bibinfo {author} {\bibfnamefont {X.}~\bibnamefont {Dong}}, \bibinfo {author} {\bibfnamefont {M.}~\bibnamefont {Dykman}},\ and\ \bibinfo {author} {\bibfnamefont {H.}~\bibnamefont {Chan}},\ }\bibfield  {title} {\bibinfo {title} {Frequency stabilization of self-sustained oscillations in a sideband-driven electromechanical resonator},\ }\href {https://doi.org/10.1103/PhysRevApplied.22.034072} {\bibfield  {journal} {\bibinfo  {journal} {Phys. Rev. Appl.}\ }\textbf {\bibinfo {volume} {22}},\ \bibinfo {pages} {034072} (\bibinfo {year} {2024})}\BibitemShut {NoStop}%
\bibitem [{\citenamefont {Houri}\ \emph {et~al.}(2021)\citenamefont {Houri}, \citenamefont {Asano}, \citenamefont {Okamoto},\ and\ \citenamefont {Yamaguchi}}]{Houri2021}%
  \BibitemOpen
  \bibfield  {author} {\bibinfo {author} {\bibfnamefont {S.}~\bibnamefont {Houri}}, \bibinfo {author} {\bibfnamefont {M.}~\bibnamefont {Asano}}, \bibinfo {author} {\bibfnamefont {H.}~\bibnamefont {Okamoto}},\ and\ \bibinfo {author} {\bibfnamefont {H.}~\bibnamefont {Yamaguchi}},\ }\bibfield  {title} {\bibinfo {title} {Self-sustained libration regime in nonlinear microelectromechanical devices},\ }\href {https://doi.org/10.1103/PhysRevApplied.16.064015} {\bibfield  {journal} {\bibinfo  {journal} {Phys. Rev. Appl.}\ }\textbf {\bibinfo {volume} {16}},\ \bibinfo {pages} {064015} (\bibinfo {year} {2021})}\BibitemShut {NoStop}%
\bibitem [{\citenamefont {Czaplewski}\ \emph {et~al.}(2018)\citenamefont {Czaplewski}, \citenamefont {Chen}, \citenamefont {Lopez}, \citenamefont {Shoshani}, \citenamefont {Eriksson}, \citenamefont {Strachan},\ and\ \citenamefont {Shaw}}]{Czaplewski2018}%
  \BibitemOpen
  \bibfield  {author} {\bibinfo {author} {\bibfnamefont {D.~A.}\ \bibnamefont {Czaplewski}}, \bibinfo {author} {\bibfnamefont {C.}~\bibnamefont {Chen}}, \bibinfo {author} {\bibfnamefont {D.}~\bibnamefont {Lopez}}, \bibinfo {author} {\bibfnamefont {O.}~\bibnamefont {Shoshani}}, \bibinfo {author} {\bibfnamefont {A.~M.}\ \bibnamefont {Eriksson}}, \bibinfo {author} {\bibfnamefont {S.}~\bibnamefont {Strachan}},\ and\ \bibinfo {author} {\bibfnamefont {S.~W.}\ \bibnamefont {Shaw}},\ }\bibfield  {title} {\bibinfo {title} {{Bifurcation Generated Mechanical Frequency Comb}},\ }\href {https://doi.org/10.1103/physrevlett.121.244302} {\bibfield  {journal} {\bibinfo  {journal} {Physical Review Letters}\ }\textbf {\bibinfo {volume} {121}},\ \bibinfo {pages} {244302} (\bibinfo {year} {2018})}\BibitemShut {NoStop}%
\bibitem [{\citenamefont {Dykman}\ \emph {et~al.}(2019)\citenamefont {Dykman}, \citenamefont {Rastelli}, \citenamefont {Roukes},\ and\ \citenamefont {Weig}}]{Dykman2019}%
  \BibitemOpen
  \bibfield  {author} {\bibinfo {author} {\bibfnamefont {M.~I.}\ \bibnamefont {Dykman}}, \bibinfo {author} {\bibfnamefont {G.}~\bibnamefont {Rastelli}}, \bibinfo {author} {\bibfnamefont {M.~L.}\ \bibnamefont {Roukes}},\ and\ \bibinfo {author} {\bibfnamefont {E.~M.}\ \bibnamefont {Weig}},\ }\bibfield  {title} {\bibinfo {title} {{Resonantly Induced Friction and Frequency Combs in Driven Nanomechanical Systems}},\ }\href {https://doi.org/10.1103/physrevlett.122.254301} {\bibfield  {journal} {\bibinfo  {journal} {Physical Review Letters}\ }\textbf {\bibinfo {volume} {122}},\ \bibinfo {pages} {254301} (\bibinfo {year} {2019})},\ \Eprint {https://arxiv.org/abs/1903.08602} {1903.08602} \BibitemShut {NoStop}%
\bibitem [{\citenamefont {Ochs}\ \emph {et~al.}(2022)\citenamefont {Ochs}, \citenamefont {Boneß}, \citenamefont {Rastelli}, \citenamefont {Seitner}, \citenamefont {Belzig}, \citenamefont {Dykman},\ and\ \citenamefont {Weig}}]{Ochs2022}%
  \BibitemOpen
  \bibfield  {author} {\bibinfo {author} {\bibfnamefont {J.~S.}\ \bibnamefont {Ochs}}, \bibinfo {author} {\bibfnamefont {D.~K.~J.}\ \bibnamefont {Boneß}}, \bibinfo {author} {\bibfnamefont {G.}~\bibnamefont {Rastelli}}, \bibinfo {author} {\bibfnamefont {M.}~\bibnamefont {Seitner}}, \bibinfo {author} {\bibfnamefont {W.}~\bibnamefont {Belzig}}, \bibinfo {author} {\bibfnamefont {M.~I.}\ \bibnamefont {Dykman}},\ and\ \bibinfo {author} {\bibfnamefont {E.~M.}\ \bibnamefont {Weig}},\ }\bibfield  {title} {\bibinfo {title} {{Frequency Comb from a Single Driven Nonlinear Nanomechanical Mode}},\ }\href {https://doi.org/10.1103/physrevx.12.041019} {\bibfield  {journal} {\bibinfo  {journal} {Physical Review X}\ }\textbf {\bibinfo {volume} {12}},\ \bibinfo {pages} {041019} (\bibinfo {year} {2022})}\BibitemShut {NoStop}%
\bibitem [{\citenamefont {Keşkekler}\ \emph {et~al.}(2022)\citenamefont {Keşkekler}, \citenamefont {Arjmandi-Tash}, \citenamefont {Steeneken},\ and\ \citenamefont {Alijani}}]{keskekler2022}%
  \BibitemOpen
  \bibfield  {author} {\bibinfo {author} {\bibfnamefont {A.}~\bibnamefont {Keşkekler}}, \bibinfo {author} {\bibfnamefont {H.}~\bibnamefont {Arjmandi-Tash}}, \bibinfo {author} {\bibfnamefont {P.~G.}\ \bibnamefont {Steeneken}},\ and\ \bibinfo {author} {\bibfnamefont {F.}~\bibnamefont {Alijani}},\ }\bibfield  {title} {\bibinfo {title} {{Symmetry-Breaking-Induced Frequency Combs in Graphene Resonators}},\ }\href {https://doi.org/10.1021/acs.nanolett.2c00360} {\bibfield  {journal} {\bibinfo  {journal} {Nano Letters}\ }\textbf {\bibinfo {volume} {22}},\ \bibinfo {pages} {6048} (\bibinfo {year} {2022})},\ \Eprint {https://arxiv.org/abs/2202.03318} {2202.03318} \BibitemShut {NoStop}%
\bibitem [{\citenamefont {Jong}\ \emph {et~al.}(2023)\citenamefont {Jong}, \citenamefont {Ganesan}, \citenamefont {Cupertino}, \citenamefont {Gröblacher},\ and\ \citenamefont {Norte}}]{Jong2023}%
  \BibitemOpen
  \bibfield  {author} {\bibinfo {author} {\bibfnamefont {M.~H. J.~d.}\ \bibnamefont {Jong}}, \bibinfo {author} {\bibfnamefont {A.}~\bibnamefont {Ganesan}}, \bibinfo {author} {\bibfnamefont {A.}~\bibnamefont {Cupertino}}, \bibinfo {author} {\bibfnamefont {S.}~\bibnamefont {Gröblacher}},\ and\ \bibinfo {author} {\bibfnamefont {R.~A.}\ \bibnamefont {Norte}},\ }\bibfield  {title} {\bibinfo {title} {{Mechanical overtone frequency combs}},\ }\href {https://doi.org/10.1038/s41467-023-36953-8} {\bibfield  {journal} {\bibinfo  {journal} {Nature Communications}\ }\textbf {\bibinfo {volume} {14}},\ \bibinfo {pages} {1458} (\bibinfo {year} {2023})},\ \Eprint {https://arxiv.org/abs/2207.06401} {2207.06401} \BibitemShut {NoStop}%
\bibitem [{\citenamefont {Wang}\ \emph {et~al.}(2022{\natexlab{b}})\citenamefont {Wang}, \citenamefont {Yang}, \citenamefont {Huan}, \citenamefont {Shi}, \citenamefont {Zhu}, \citenamefont {Jiang}, \citenamefont {Deng},\ and\ \citenamefont {Wei}}]{Wang2022comb}%
  \BibitemOpen
  \bibfield  {author} {\bibinfo {author} {\bibfnamefont {X.}~\bibnamefont {Wang}}, \bibinfo {author} {\bibfnamefont {Q.}~\bibnamefont {Yang}}, \bibinfo {author} {\bibfnamefont {R.}~\bibnamefont {Huan}}, \bibinfo {author} {\bibfnamefont {Z.}~\bibnamefont {Shi}}, \bibinfo {author} {\bibfnamefont {W.}~\bibnamefont {Zhu}}, \bibinfo {author} {\bibfnamefont {Z.}~\bibnamefont {Jiang}}, \bibinfo {author} {\bibfnamefont {Z.}~\bibnamefont {Deng}},\ and\ \bibinfo {author} {\bibfnamefont {X.}~\bibnamefont {Wei}},\ }\bibfield  {title} {\bibinfo {title} {{Frequency comb in 1:3 internal resonance of coupled micromechanical resonators}},\ }\href {https://doi.org/10.1063/5.0091237} {\bibfield  {journal} {\bibinfo  {journal} {Applied Physics Letters}\ }\textbf {\bibinfo {volume} {120}},\ \bibinfo {pages} {173506} (\bibinfo {year} {2022}{\natexlab{b}})}\BibitemShut {NoStop}%
\bibitem [{\citenamefont {Singh}\ \emph {et~al.}(2020)\citenamefont {Singh}, \citenamefont {Sarkar}, \citenamefont {Guria}, \citenamefont {Nicholl}, \citenamefont {Chakraborty}, \citenamefont {Bolotin},\ and\ \citenamefont {Ghosh}}]{Singh2020}%
  \BibitemOpen
  \bibfield  {author} {\bibinfo {author} {\bibfnamefont {R.}~\bibnamefont {Singh}}, \bibinfo {author} {\bibfnamefont {A.}~\bibnamefont {Sarkar}}, \bibinfo {author} {\bibfnamefont {C.}~\bibnamefont {Guria}}, \bibinfo {author} {\bibfnamefont {R.~J.}\ \bibnamefont {Nicholl}}, \bibinfo {author} {\bibfnamefont {S.}~\bibnamefont {Chakraborty}}, \bibinfo {author} {\bibfnamefont {K.~I.}\ \bibnamefont {Bolotin}},\ and\ \bibinfo {author} {\bibfnamefont {S.}~\bibnamefont {Ghosh}},\ }\bibfield  {title} {\bibinfo {title} {{Giant Tunable Mechanical Nonlinearity in Graphene–Silicon Nitride Hybrid Resonator}},\ }\href {https://doi.org/10.1021/acs.nanolett.0c01586} {\bibfield  {journal} {\bibinfo  {journal} {Nano Letters}\ }\textbf {\bibinfo {volume} {20}},\ \bibinfo {pages} {4659} (\bibinfo {year} {2020})},\ \Eprint {https://arxiv.org/abs/1904.01613} {1904.01613} \BibitemShut {NoStop}%
\bibitem [{\citenamefont {Mouharrar}\ \emph {et~al.}(2024)\citenamefont {Mouharrar}, \citenamefont {Rahmanian}, \citenamefont {Abdelrahman}, \citenamefont {Shama}, \citenamefont {Akbari}, \citenamefont {Yavuz},\ and\ \citenamefont {Abdel-Rahman}}]{Mouharrar2024}%
  \BibitemOpen
  \bibfield  {author} {\bibinfo {author} {\bibfnamefont {H.}~\bibnamefont {Mouharrar}}, \bibinfo {author} {\bibfnamefont {S.}~\bibnamefont {Rahmanian}}, \bibinfo {author} {\bibfnamefont {R.}~\bibnamefont {Abdelrahman}}, \bibinfo {author} {\bibfnamefont {Y.~S.}\ \bibnamefont {Shama}}, \bibinfo {author} {\bibfnamefont {M.}~\bibnamefont {Akbari}}, \bibinfo {author} {\bibfnamefont {M.}~\bibnamefont {Yavuz}},\ and\ \bibinfo {author} {\bibfnamefont {E.}~\bibnamefont {Abdel-Rahman}},\ }\bibfield  {title} {\bibinfo {title} {{Generation of Soliton Frequency Combs in NEMS}},\ }\bibfield  {journal} {\bibinfo  {journal} {Nano Letters}\ }\href {https://doi.org/10.1021/acs.nanolett.4c02249} {10.1021/acs.nanolett.4c02249} (\bibinfo {year} {2024})\BibitemShut {NoStop}%
\bibitem [{\citenamefont {Houri}\ \emph {et~al.}(2020)\citenamefont {Houri}, \citenamefont {Asano}, \citenamefont {Yamaguchi}, \citenamefont {Yoshimura}, \citenamefont {Koike},\ and\ \citenamefont {Minati}}]{Houri2020}%
  \BibitemOpen
  \bibfield  {author} {\bibinfo {author} {\bibfnamefont {S.}~\bibnamefont {Houri}}, \bibinfo {author} {\bibfnamefont {M.}~\bibnamefont {Asano}}, \bibinfo {author} {\bibfnamefont {H.}~\bibnamefont {Yamaguchi}}, \bibinfo {author} {\bibfnamefont {N.}~\bibnamefont {Yoshimura}}, \bibinfo {author} {\bibfnamefont {Y.}~\bibnamefont {Koike}},\ and\ \bibinfo {author} {\bibfnamefont {L.}~\bibnamefont {Minati}},\ }\bibfield  {title} {\bibinfo {title} {{Generic Rotating-Frame-Based Approach to Chaos Generation in Nonlinear Micro- and Nanoelectromechanical System Resonators}},\ }\href {https://doi.org/10.1103/physrevlett.125.174301} {\bibfield  {journal} {\bibinfo  {journal} {Physical Review Letters}\ }\textbf {\bibinfo {volume} {125}},\ \bibinfo {pages} {174301} (\bibinfo {year} {2020})},\ \Eprint {https://arxiv.org/abs/2005.09215} {2005.09215} \BibitemShut {NoStop}%
\bibitem [{\citenamefont {Defoort}\ \emph {et~al.}(2021)\citenamefont {Defoort}, \citenamefont {Rufer}, \citenamefont {Fesquet},\ and\ \citenamefont {Basrour}}]{Defoort2021}%
  \BibitemOpen
  \bibfield  {author} {\bibinfo {author} {\bibfnamefont {M.}~\bibnamefont {Defoort}}, \bibinfo {author} {\bibfnamefont {L.}~\bibnamefont {Rufer}}, \bibinfo {author} {\bibfnamefont {L.}~\bibnamefont {Fesquet}},\ and\ \bibinfo {author} {\bibfnamefont {S.}~\bibnamefont {Basrour}},\ }\bibfield  {title} {\bibinfo {title} {{A dynamical approach to generate chaos in a micromechanical resonator}},\ }\href {https://doi.org/10.1038/s41378-021-00241-6} {\bibfield  {journal} {\bibinfo  {journal} {Microsystems \& Nanoengineering}\ }\textbf {\bibinfo {volume} {7}},\ \bibinfo {pages} {17} (\bibinfo {year} {2021})},\ \Eprint {https://arxiv.org/abs/2101.11976} {2101.11976} \BibitemShut {NoStop}%
\bibitem [{\citenamefont {Shoshani}\ and\ \citenamefont {Shaw}(2023)}]{Shoshani2023}%
  \BibitemOpen
  \bibfield  {author} {\bibinfo {author} {\bibfnamefont {O.}~\bibnamefont {Shoshani}}\ and\ \bibinfo {author} {\bibfnamefont {S.~W.}\ \bibnamefont {Shaw}},\ }\bibfield  {title} {\bibinfo {title} {{Nonlinear interactions between vibration modes with vastly different eigenfrequencies}},\ }\href {https://doi.org/10.1038/s42005-023-01323-9} {\bibfield  {journal} {\bibinfo  {journal} {Communications Physics}\ }\textbf {\bibinfo {volume} {6}},\ \bibinfo {pages} {213} (\bibinfo {year} {2023})}\BibitemShut {NoStop}%
\bibitem [{\citenamefont {Eriksson}\ \emph {et~al.}(2023)\citenamefont {Eriksson}, \citenamefont {Shoshani}, \citenamefont {López}, \citenamefont {Shaw},\ and\ \citenamefont {Czaplewski}}]{Eriksson2023}%
  \BibitemOpen
  \bibfield  {author} {\bibinfo {author} {\bibfnamefont {A.~M.}\ \bibnamefont {Eriksson}}, \bibinfo {author} {\bibfnamefont {O.}~\bibnamefont {Shoshani}}, \bibinfo {author} {\bibfnamefont {D.}~\bibnamefont {López}}, \bibinfo {author} {\bibfnamefont {S.~W.}\ \bibnamefont {Shaw}},\ and\ \bibinfo {author} {\bibfnamefont {D.~A.}\ \bibnamefont {Czaplewski}},\ }\bibfield  {title} {\bibinfo {title} {{Controllable branching of robust response patterns in nonlinear mechanical resonators}},\ }\href {https://doi.org/10.1038/s41467-022-35685-5} {\bibfield  {journal} {\bibinfo  {journal} {Nature Communications}\ }\textbf {\bibinfo {volume} {14}},\ \bibinfo {pages} {161} (\bibinfo {year} {2023})}\BibitemShut {NoStop}%
\bibitem [{\citenamefont {Houri}\ \emph {et~al.}(2019)\citenamefont {Houri}, \citenamefont {Hatanaka}, \citenamefont {Asano}, \citenamefont {Ohta},\ and\ \citenamefont {Yamaguchi}}]{Houri2019}%
  \BibitemOpen
  \bibfield  {author} {\bibinfo {author} {\bibfnamefont {S.}~\bibnamefont {Houri}}, \bibinfo {author} {\bibfnamefont {D.}~\bibnamefont {Hatanaka}}, \bibinfo {author} {\bibfnamefont {M.}~\bibnamefont {Asano}}, \bibinfo {author} {\bibfnamefont {R.}~\bibnamefont {Ohta}},\ and\ \bibinfo {author} {\bibfnamefont {H.}~\bibnamefont {Yamaguchi}},\ }\bibfield  {title} {\bibinfo {title} {{Limit cycles and bifurcations in a nonlinear MEMS resonator with a 1:3 internal resonance}},\ }\href {https://doi.org/10.1063/1.5085219} {\bibfield  {journal} {\bibinfo  {journal} {Applied Physics Letters}\ }\textbf {\bibinfo {volume} {114}},\ \bibinfo {pages} {103103} (\bibinfo {year} {2019})}\BibitemShut {NoStop}%
\bibitem [{\citenamefont {Luo}\ \emph {et~al.}(2021)\citenamefont {Luo}, \citenamefont {Gao},\ and\ \citenamefont {Liu}}]{Luo2021}%
  \BibitemOpen
  \bibfield  {author} {\bibinfo {author} {\bibfnamefont {W.}~\bibnamefont {Luo}}, \bibinfo {author} {\bibfnamefont {N.}~\bibnamefont {Gao}},\ and\ \bibinfo {author} {\bibfnamefont {D.}~\bibnamefont {Liu}},\ }\bibfield  {title} {\bibinfo {title} {{Multimode Nonlinear Coupling Induced by Internal Resonance in a Microcantilever Resonator}},\ }\href {https://doi.org/10.1021/acs.nanolett.0c04301} {\bibfield  {journal} {\bibinfo  {journal} {Nano Letters}\ }\textbf {\bibinfo {volume} {21}},\ \bibinfo {pages} {1062} (\bibinfo {year} {2021})}\BibitemShut {NoStop}%
\bibitem [{\citenamefont {Czaplewski}\ \emph {et~al.}(2019)\citenamefont {Czaplewski}, \citenamefont {Strachan}, \citenamefont {Shoshani}, \citenamefont {Shaw},\ and\ \citenamefont {López}}]{Czaplewski2019}%
  \BibitemOpen
  \bibfield  {author} {\bibinfo {author} {\bibfnamefont {D.~A.}\ \bibnamefont {Czaplewski}}, \bibinfo {author} {\bibfnamefont {S.}~\bibnamefont {Strachan}}, \bibinfo {author} {\bibfnamefont {O.}~\bibnamefont {Shoshani}}, \bibinfo {author} {\bibfnamefont {S.~W.}\ \bibnamefont {Shaw}},\ and\ \bibinfo {author} {\bibfnamefont {D.}~\bibnamefont {López}},\ }\bibfield  {title} {\bibinfo {title} {{Bifurcation diagram and dynamic response of a MEMS resonator with a 1:3 internal resonance}},\ }\href {https://doi.org/10.1063/1.5099459} {\bibfield  {journal} {\bibinfo  {journal} {Applied Physics Letters}\ }\textbf {\bibinfo {volume} {114}},\ \bibinfo {pages} {254104} (\bibinfo {year} {2019})}\BibitemShut {NoStop}%
\bibitem [{\citenamefont {Bückle}\ \emph {et~al.}(2021)\citenamefont {Bückle}, \citenamefont {Klaß}, \citenamefont {Nägele}, \citenamefont {Braive},\ and\ \citenamefont {Weig}}]{Bückle2021}%
  \BibitemOpen
  \bibfield  {author} {\bibinfo {author} {\bibfnamefont {M.}~\bibnamefont {Bückle}}, \bibinfo {author} {\bibfnamefont {Y.~S.}\ \bibnamefont {Klaß}}, \bibinfo {author} {\bibfnamefont {F.~B.}\ \bibnamefont {Nägele}}, \bibinfo {author} {\bibfnamefont {R.}~\bibnamefont {Braive}},\ and\ \bibinfo {author} {\bibfnamefont {E.~M.}\ \bibnamefont {Weig}},\ }\bibfield  {title} {\bibinfo {title} {{Universal Length Dependence of Tensile Stress in Nanomechanical String Resonators}},\ }\href {https://doi.org/10.1103/physrevapplied.15.034063} {\bibfield  {journal} {\bibinfo  {journal} {Physical Review Applied}\ }\textbf {\bibinfo {volume} {15}},\ \bibinfo {pages} {034063} (\bibinfo {year} {2021})},\ \Eprint {https://arxiv.org/abs/2012.00507} {2012.00507} \BibitemShut {NoStop}%
\bibitem [{\citenamefont {Klaß}\ \emph {et~al.}(2022)\citenamefont {Klaß}, \citenamefont {Doster}, \citenamefont {Bückle}, \citenamefont {Braive},\ and\ \citenamefont {Weig}}]{Klaß2022}%
  \BibitemOpen
  \bibfield  {author} {\bibinfo {author} {\bibfnamefont {Y.~S.}\ \bibnamefont {Klaß}}, \bibinfo {author} {\bibfnamefont {J.}~\bibnamefont {Doster}}, \bibinfo {author} {\bibfnamefont {M.}~\bibnamefont {Bückle}}, \bibinfo {author} {\bibfnamefont {R.}~\bibnamefont {Braive}},\ and\ \bibinfo {author} {\bibfnamefont {E.~M.}\ \bibnamefont {Weig}},\ }\bibfield  {title} {\bibinfo {title} {Determining young's modulus via the eigenmode spectrum of a nanomechanical string resonator},\ }\href {https://doi.org/10.1063/5.0100405} {\bibfield  {journal} {\bibinfo  {journal} {Applied Physics Letters}\ }\textbf {\bibinfo {volume} {121}},\ \bibinfo {pages} {083501} (\bibinfo {year} {2022})}\BibitemShut {NoStop}%
\bibitem [{\citenamefont {Landau}\ and\ \citenamefont {Lifshitz}(1976)}]{landau1976}%
  \BibitemOpen
  \bibfield  {author} {\bibinfo {author} {\bibfnamefont {L.~D.}\ \bibnamefont {Landau}}\ and\ \bibinfo {author} {\bibfnamefont {E.~M.}\ \bibnamefont {Lifshitz}},\ }\href@noop {} {\emph {\bibinfo {title} {Mechanics}}},\ \bibinfo {edition} {2nd}\ ed.,\ Vol.~\bibinfo {volume} {1}\ (\bibinfo  {publisher} {Pergamon Press},\ \bibinfo {address} {Oxford},\ \bibinfo {year} {1976})\BibitemShut {NoStop}%
\bibitem [{\citenamefont {Ochs}\ \emph {et~al.}(2021{\natexlab{b}})\citenamefont {Ochs}, \citenamefont {Rastelli}, \citenamefont {Seitner}, \citenamefont {Dykman},\ and\ \citenamefont {Weig}}]{Ochs2021resonant}%
  \BibitemOpen
  \bibfield  {author} {\bibinfo {author} {\bibfnamefont {J.~S.}\ \bibnamefont {Ochs}}, \bibinfo {author} {\bibfnamefont {G.}~\bibnamefont {Rastelli}}, \bibinfo {author} {\bibfnamefont {M.}~\bibnamefont {Seitner}}, \bibinfo {author} {\bibfnamefont {M.~I.}\ \bibnamefont {Dykman}},\ and\ \bibinfo {author} {\bibfnamefont {E.~M.}\ \bibnamefont {Weig}},\ }\bibfield  {title} {\bibinfo {title} {Resonant nonlinear response of a nanomechanical system with broken symmetry},\ }\href {https://doi.org/10.1103/PhysRevB.104.155434} {\bibfield  {journal} {\bibinfo  {journal} {Phys. Rev. B}\ }\textbf {\bibinfo {volume} {104}},\ \bibinfo {pages} {155434} (\bibinfo {year} {2021}{\natexlab{b}})}\BibitemShut {NoStop}%
\bibitem [{\citenamefont {Barquist}\ \emph {et~al.}(2020)\citenamefont {Barquist}, \citenamefont {Jiang}, \citenamefont {Gunther}, \citenamefont {Eng}, \citenamefont {Lee},\ and\ \citenamefont {Chan}}]{Barquist2020}%
  \BibitemOpen
  \bibfield  {author} {\bibinfo {author} {\bibfnamefont {C.~S.}\ \bibnamefont {Barquist}}, \bibinfo {author} {\bibfnamefont {W.~G.}\ \bibnamefont {Jiang}}, \bibinfo {author} {\bibfnamefont {K.}~\bibnamefont {Gunther}}, \bibinfo {author} {\bibfnamefont {N.}~\bibnamefont {Eng}}, \bibinfo {author} {\bibfnamefont {Y.}~\bibnamefont {Lee}},\ and\ \bibinfo {author} {\bibfnamefont {H.~B.}\ \bibnamefont {Chan}},\ }\bibfield  {title} {\bibinfo {title} {Damping of a microelectromechanical oscillator in turbulent superfluid $^{4}\mathrm{He}$: A probe of quantized vorticity in the ultralow temperature regime},\ }\href {https://doi.org/10.1103/PhysRevB.101.174513} {\bibfield  {journal} {\bibinfo  {journal} {Phys. Rev. B}\ }\textbf {\bibinfo {volume} {101}},\ \bibinfo {pages} {174513} (\bibinfo {year} {2020})}\BibitemShut {NoStop}%
\bibitem [{\citenamefont {Guthrie}\ \emph {et~al.}(2021)\citenamefont {Guthrie}, \citenamefont {Kafanov}, \citenamefont {Noble}, \citenamefont {Pashkin}, \citenamefont {Pickett}, \citenamefont {Tsepelin}, \citenamefont {Dorofeev}, \citenamefont {Krupenin},\ and\ \citenamefont {Presnov}}]{Guthrie2021}%
  \BibitemOpen
  \bibfield  {author} {\bibinfo {author} {\bibfnamefont {A.}~\bibnamefont {Guthrie}}, \bibinfo {author} {\bibfnamefont {S.}~\bibnamefont {Kafanov}}, \bibinfo {author} {\bibfnamefont {M.~T.}\ \bibnamefont {Noble}}, \bibinfo {author} {\bibfnamefont {Y.~A.}\ \bibnamefont {Pashkin}}, \bibinfo {author} {\bibfnamefont {G.~R.}\ \bibnamefont {Pickett}}, \bibinfo {author} {\bibfnamefont {V.}~\bibnamefont {Tsepelin}}, \bibinfo {author} {\bibfnamefont {A.~A.}\ \bibnamefont {Dorofeev}}, \bibinfo {author} {\bibfnamefont {V.~A.}\ \bibnamefont {Krupenin}},\ and\ \bibinfo {author} {\bibfnamefont {D.~E.}\ \bibnamefont {Presnov}},\ }\bibfield  {title} {\bibinfo {title} {{Nanoscale real-time detection of quantum vortices at millikelvin temperatures}},\ }\href {https://doi.org/10.1038/s41467-021-22909-3} {\bibfield  {journal} {\bibinfo  {journal} {Nature Communications}\ }\textbf {\bibinfo {volume} {12}},\ \bibinfo {pages} {2645} (\bibinfo {year} {2021})}\BibitemShut {NoStop}%
\bibitem [{\citenamefont {Barquist}\ \emph {et~al.}(2022)\citenamefont {Barquist}, \citenamefont {Jiang}, \citenamefont {Gunther}, \citenamefont {Lee},\ and\ \citenamefont {Chan}}]{Barquist2022}%
  \BibitemOpen
  \bibfield  {author} {\bibinfo {author} {\bibfnamefont {C.~S.}\ \bibnamefont {Barquist}}, \bibinfo {author} {\bibfnamefont {W.~G.}\ \bibnamefont {Jiang}}, \bibinfo {author} {\bibfnamefont {K.}~\bibnamefont {Gunther}}, \bibinfo {author} {\bibfnamefont {Y.}~\bibnamefont {Lee}},\ and\ \bibinfo {author} {\bibfnamefont {H.~B.}\ \bibnamefont {Chan}},\ }\bibfield  {title} {\bibinfo {title} {Phase noise in a duffing oscillator induced by quantum turbulence},\ }\href {https://doi.org/10.1103/PhysRevB.106.094502} {\bibfield  {journal} {\bibinfo  {journal} {Phys. Rev. B}\ }\textbf {\bibinfo {volume} {106}},\ \bibinfo {pages} {094502} (\bibinfo {year} {2022})}\BibitemShut {NoStop}%
\end{thebibliography}%

\end{document}


\title{Supplemental Material:\\
Non-Exponential Relaxation in the Rotating Frame of a Driven Nanomechanical Mode}

\author{Hyunjin Choi}
\author{Oriel Shoshani}
\author{Ryundon Kim}
\author{Younghun Ryu}
\author{Jinhoon Jeong}
\author{Junho Suh}
\author{Steven W. Shaw}
\author{M. I. Dykman}
\author{Hyoungsoon Choi}

\maketitle

\section{Theory}
The equations of motion in the rotating frame, $\dot{q}=\partial_{p} \mathcal{H} -\Gamma q, \; \dot{p}=-\partial_{q}\mathcal{H}-\Gamma p$ is obtained by adding linear damping to the Hamiltonian $\mathcal{H}_0$ and $\mathcal{H}_1$ given in the main text with
\begin{align*}
&\mathcal{H}_0(q,p)=\frac{3\gamma}{32\omega_1}(q^{2}\!+\!p^{2})^2 - \frac{1}{2}\delta \omega_1(q^2\!+\!p^2) -\frac{F_{1}q}{2\omega_{1}},\\
&\mathcal{H}_1(q,p,t)= - {F_{2}}[q\cos(\delta\omega_2 t)-p\sin(\delta\omega_2 t)]/2\omega_1.
\end{align*}
The equations are then written explicitly as
\begin{align}
\dot q&=-\left(\delta\omega_1-\frac{3\gamma}{8\omega_1}(q^2+p^2)\right)p-\frac{F_2}{2\omega_1}\sin(\delta\omega_{2}t)-\Gamma q,
\label{eq:q}\\
\dot p&=\left(\delta\omega_1-\frac{3\gamma}{8\omega_1}(q^2+p^2)\right)q+\frac{F_1}{2\omega_1}-\frac{ F_2}{2\omega_1}\cos(\delta\omega_{2}t)-\Gamma p.
\label{eq:p}
\end{align}
The conservative dynamics ($\Gamma=0$, $F_2=0$) can be viewed as the motion of a ''particle'' with position $q$, "momentum" $p$ and Hamiltonian $\mathcal{H}_0$ . Therefore, when $\gamma>0$ and $\delta\omega_1>0$, the extrema of $\mathcal{H}_0$ correspond
to the equilibrium points of the particle for which $p=0$, and $q=q_{\rm eq}$. These equilibrium points are found by solving the algebraic equation
\begin{align}
\frac{F_1}{2\omega_1}-\frac{3\gamma}{8\omega_1}q_{\rm eq}^3+\delta\omega_1 q_{\rm eq}=0.
\label{eq:q_eq}
\end{align}
For $F_1>F_{\rm{cr_1}}$, this equation has three real solutions. A straightforward linear stability analysis shows that the determinant of the corresponding Jacobian is given by $\Delta=\partial^2_q\mathcal{H}_0|_{(q_{\rm eq},0)}\partial^2_p\mathcal{H}_0|_{(q_{\rm eq},0)}$, which is positive for the intermediate and largest solutions of Eq.~\eqref{eq:q_eq}, and negative for the smallest solution. Therefore, the intermediate and largest equilibrium points are centers, and the smallest equilibrium point is a saddle. We focus here on cases where the inclusion of $\mathcal{H}_1$ ($F_2 \neq 0$) and the light damping $\Gamma$ lead to oscillations in the vicinity of the stable high-amplitude state $(a_{\rm hi},0)$ where $a_{\rm hi}$ can be approximated by
\begin{align}
a_{\rm hi}\equiv q_{\rm eq}\approx\sqrt{\frac{8\omega_1\delta\omega_1}{3\gamma}+F_1\sqrt{\frac{2}{3\gamma\omega_1\delta\omega_1}}}.\label{eq:qeq}
\end{align}
The amplitude approximation in Eq.~\eqref{eq:qeq} is shown in Fig.~1(b) of the main text and agrees well with the experimental measurements.

To analyze the oscillations around $(q_{\rm eq},0)$, we introduce the local coordinate $u=q-q_{\rm eq}$, which yields the following dynamical system for the deviation away from the high-amplitude state
\begin{align}
\dot u&=\left(\frac{3\gamma q^2_{\rm eq}}{8\omega_1}-\delta\omega_1\right)p+\frac{3\gamma}{8\omega_1}\left[p^2+u(u+2q_{\rm eq})\right]p+\frac{F_2}{2\omega_{1}}\sin(\delta\omega_2 t)-\Gamma u,\label{eq:u}\\
\dot p&=-\left(\frac{9\gamma q^2_{\rm eq}}{8\omega_1}-\delta\omega_1\right)u-\frac{3\gamma}{8\omega_1}\left[p^2(q_{\rm eq}-u)+u^2(u+3q_{\rm eq})\right]+\frac{F_2}{2\omega_{1}}\cos(\delta\omega_2 t)-\Gamma p.\label{eq:pdev}
\end{align}
The eigenfrequency of small deviations ($|u|,|p|\ll q_{\rm eq}$) is given by
$$\Omega=\sqrt{\left(\frac{3\gamma q^2_{\rm eq}}{8\omega_1}-\delta\omega_1\right)\left(\frac{9\gamma q^2_{\rm eq}}{8\omega_1}-\delta\omega_1\right)}=\sqrt{\frac{F_1}{32\omega_1^2}\left(\frac{9\gamma F_1}{\omega_1\delta\omega_1}+8\sqrt{6\gamma\omega_1\delta\omega_1}\right)}.$$
For a near-resonant secondary drive $|\delta\omega_2-\Omega|/\Omega\ll1$, $u$ and $p$ can oscillate non-linearly with large amplitude around $(q_{\rm eq},0)$. When the secondary drive is weak, $F_2\ll F_1$, we can apply a standard perturbation analysis of weakly nonlinear oscillators. To this end, we introduce a bookkeeping parameter $\epsilon$ to keep track of small quantities in Eqs.~ \eqref{eq:u}-\eqref{eq:pdev}
\begin{align}
\dot u&=\left(\frac{3\gamma q^2_{\rm eq}}{8\omega_1}-\delta\omega_1\right)p+\frac{3\gamma}{8\omega_1}\left[p^2+u(u+2q_{\rm eq})\right]p+\frac{\epsilon^3F_2}{2\omega_{1}}\sin(\delta\omega_2 t)-\epsilon^2\Gamma u,\label{eq:urs}\\
\dot p&=-\left(\frac{9\gamma q^2_{\rm eq}}{8\omega_1}-\delta\omega_1\right)u-\frac{3\gamma}{8\omega_1}\left[p^2(q_{\rm eq}\!-\!u)\!+\!u^2(u\!+\!3q_{\rm eq})\right]+\frac{\epsilon^3F_2}{2\omega_{1}}\cos(\delta\omega_2 t)-\epsilon^2\Gamma p,\label{eq:pdevrs}
\end{align}
and use the asymptotic expansion
$$u(t)=\epsilon u^{(1)}(t,\epsilon^2 t)+\epsilon^2 u^{(2)}(t,\epsilon^2 t)+\epsilon^3 u^{(3)}(t,\epsilon^2 t)+{\mathcal{O}}(\epsilon^4),~p(t)=\epsilon p^{(1)}(t,\epsilon^2 t)+\epsilon^2 p^{(2)}(t,\epsilon^2 t)+\epsilon^3 p^{(3)}(t,\epsilon^2 t)+{\mathcal{O}}(\epsilon^4),
$$
where we explicitly denote the presence of a slow time-scale $\epsilon^2 t$ that is associated with the relaxation time of the system $1/(\epsilon^2\Gamma )$. By substituting the asymptotic expansion into Eqs.~\eqref{eq:urs}-\eqref{eq:pdevrs}, rearranging them, and equating the same power of $\epsilon$ to zero, we obtain the following set of equations:
\begin{align}
\mathcal{O}(\epsilon):~\ddot u^{(1)}+\Omega^2u^{(1)}&=0, \label{eq:u1}\\
\mathcal{O}(\epsilon^2):~\ddot u^{(2)}+\Omega^2u^{(2)}&=\frac{3\gamma q_{\rm eq}}{64\omega_1^2}\left[ (8\omega_1\delta\omega_1 - 3\gamma q_{\rm eq}^2)(3u^{(1)2}+p^{(1)2})+16\omega_1(u^{(1)}\dot p^{(1)} +\dot u^{(1)} p^{(1)})\right], \label{eq:u2}\\
\mathcal{O}(\epsilon^3):~\ddot u^{(3)}+\Omega^2u^{(3)}&=\frac{F_2[\omega_1(\delta\omega_2-\delta\omega_1)+3\gamma q_{\rm eq}^2]}{16\omega_1^2}\cos(\delta\omega_2 t)-2\partial_{\epsilon^2t}\dot u^{(1)}-\Gamma\left[\dot u^{(1)}+\left(\frac{3\gamma q^2_{\rm eq}}{8\omega_1}-\delta\omega_1\right)p^{(1)}\right]\nonumber\\
&+\frac{3\gamma q_{\rm eq}}{4\omega_1}\left[u^{(2)}\dot p^{(1)}+\dot u^{(2)} p^{(1)}+u^{(1)}\dot p^{(2)}+\dot u^{(1)} p^{(2)}-\left(\frac{3\gamma q^2_{\rm eq}}{8\omega_1}-\delta\omega_1\right)(3u^{(1)}u^{(2)}+p^{(1)}p^{(2)})\right]\nonumber\\
& +\frac{3\gamma}{8\omega_1}\left[\dot p^{(1)}(3p^{(1)2}+u^{(1)2})+2u^{(1)}p^{(1)}\dot u^{(1)} +\left(\frac{3\gamma q^2_{\rm eq}}{8\omega_1}-\delta\omega_1\right)u^{(1)}(p^{(1)2}-u^{(1)2})\right],\label{eq:u3}
\end{align}
where $p^{(j)}$ can be computed from the following expressions
\begin{align}
&\left(\frac{3\gamma q^2_{\rm eq}}{8\omega_1}-\delta\omega_1\right)p^{(1)}=\dot u^{(1)},~~~~~~~\left(\frac{3\gamma q^2_{\rm eq}}{8\omega_1}-\delta\omega_1\right)p^{(2)}=\dot u^{(2)}-\frac{3\gamma q_{\rm eq}}{4\omega_1}p^{(1)}u^{(1)},\nonumber\\
&\left(\frac{3\gamma q^2_{\rm eq}}{8\omega_1}-\delta\omega_1\right)p^{(3)}=\dot u^{(3)}+\Gamma u^{(1)}-\frac{F_2}{2\omega_1}\sin(\delta\omega_2 t)+\frac{3\gamma}{8\omega_1}\left[2q_{\rm eq}(u^{(1)}p^{(2)}+u^{(2)}p^{(1)})+p^{(1)}(p^{(1)2}+u^{(1)2}) \right]. \nonumber
\end{align}
The solution of Eq.~\eqref{eq:u1} is given by $u^{(1)}(t,\epsilon^2t)=a(\epsilon^2t)\cos(\Omega t+\phi(\epsilon^2t))$, which can be substituted into Eqs.~\eqref{eq:u2}-\eqref{eq:u3} to determine $u^{(2)}$, $u^{(3)}$, and a pair of slow evolution equations for $a(\epsilon^2t)$ and $\phi(\epsilon^2t)$ [from the fundamental harmonic components in Eq.~\eqref{eq:u3}]. By defining a modified phase $\theta=\phi-(\delta\omega_2-\Omega)t$, which implies a resonant secondary drive $|\delta\omega_2-\Omega|\sim \mathcal{O}(\epsilon^2)$, we arrive to the following slow evolution equations
\begin{align}
\frac{da}{d(\epsilon^2t)}&=-\Gamma a-\frac{F_{\rm 2eff}}{4\omega_1}\sin\theta,\label{eq:a}\\
\frac{d\theta}{d(\epsilon^2t)}&=\Omega-\delta\omega_2+\frac{3\gamma_{\rm eff}}{8\Omega}a^2-\frac{F_{\rm 2eff}}{4\omega_1 a}\cos\theta,\label{eq:theta}
\end{align}
where
\begin{align}
\gamma_{\rm eff}=\frac{\gamma(-81\gamma^3q_{\rm eq}^6 - 504\gamma^2q_{\rm eq}^4\omega_1 \delta \omega_1 + 576 \gamma q_{\rm eq}^2\omega_1^2\delta \omega_1^2 + 512\omega_1^3\delta \omega_1^3)}{1024\omega_1^4\Omega^2}~\mathrm{and}~F_{\rm 2eff}=F_2\left(1+\sqrt{\frac{8\omega_1\delta\omega_1 - 3\gamma q^2_{\rm eq}}{8\omega_1\delta\omega_1 - 9\gamma q^2_{\rm eq}}}\right)
\end{align}

represent the effective Duffing coefficient of the response and the effective drive amplitude of the in-phase component, respectively. We note that while Eqs.~\eqref{eq:a}-\eqref{eq:theta} represent oscillations in the rotating frame around $(q_{\rm eq},0)$, their structure is similar to the structure of the phase and amplitude equations of a standard Duffing oscillator. Therefore, the same techniques can be applied to analyze these equations and we find the threshold drive amplitude $F_{\rm cr_2}$ given by
$$F_{\rm cr_2}=\frac{32}{1+\sqrt{(8\omega_1\delta\omega_1 - 3\gamma q^2_{\rm eq})/(8\omega_1\delta\omega_1 - 9\gamma q^2_{\rm eq})}}\sqrt{\Omega\omega_1^2\Gamma^3/(9\sqrt{3}|\gamma_{\rm eff}|)}$$
such that bi-stable response exists for $F_2>F_{\rm cr_2}$. With the inclusion of a DC offset and the secondary overtone, the expressions for $u$ and $p$ are given by
\begin{align}
&u=\frac{3\gamma q_{\rm eq} a^2(16\omega_1\delta\omega_1-9\gamma q_{\rm eq}^2)}{(8\omega_1\delta\omega_1 - 9\gamma q_{\rm eq}^2)(8\omega_1\delta\omega_1 - 3\gamma q_{\rm eq}^2)} + a\cos(\delta\omega_2t - \theta)\nonumber\\
&- \frac{6\gamma q_{\rm eq} a^2(4\omega_1\delta\omega_1-3\gamma q_{\rm eq}^2)}{(8\omega_1\delta\omega_1 - 9\gamma q_{\rm eq}^2)(8\omega_1\delta\omega_1 - 3\gamma q_{\rm eq}^2)}\cos(2\delta\omega_2 t-2\theta),\label{eq:u_ot}\\
&p= \frac{8\omega_1 a}{8\omega_1\delta\omega_1 - 3\gamma q_{\rm eq}^2}[\delta\omega_2\sin(\delta\omega_2t - \theta)-\Gamma\cos(\delta\omega_2t - \theta)]+\frac{4F_2}{8\omega_1\delta\omega_1 - 3\gamma q_{\rm eq}^2}\sin(\delta\omega_2t)\nonumber\\
&+\frac{24\gamma q_{\rm eq} \omega_1\delta\omega_2 a^2}{(8\omega_1\delta\omega_1 - 9\gamma q_{\rm eq})(8\omega_1\delta\omega_1 - 3\gamma q_{\rm eq})}\sin(2\delta\omega_2t-2\theta).\label{eq:p_ot}
\end{align}
Note that Eqs.~\eqref{eq:u_ot}-\eqref{eq:p_ot} can also describe the ring-down response with $F_2$ set to zero, and the frequency and amplitude replaced by $\delta\omega_2\rightarrow\Omega$ and $a\rightarrow a(0) \ee^{-\Gamma t}$. Therefore, we see that while the fundamental harmonic decays with $\ee^{-\Gamma t}$, the second harmonic and the DC offset decay with $\ee^{-2\Gamma t}$.

An alternative and more global way to describe the dynamics is by using an action-angle representation of the oscillations in the rotating frame, where $I_{\rm ho}=(q^2+p^2)/2$ is the action, and $\tan^{-1}(p/q)=\phi_{\rm ho}$ is the angle. The Hamiltonian $\mathcal{H}_0$ and the evolution equations of the action and angle are given by
\begin{align}
\mathcal{H}_0&=-\Delta\omega_1I_{\rm ho}+\frac{3\gamma}{8\omega_1}I_{\rm ho}^2-\frac{F_1}{\omega_1}\sqrt{\frac{I_{\rm ho}}{2}}\cos\phi_{\rm ho}\label{Ham_AA}
\\\dot \phi_{\rm ho}&=\frac{\partial\mathcal H_0}{\partial I_{\rm ho}}-\mathcal{F}_{\phi_{\rm ho}}\nonumber
\\ &=-\Delta\omega_1+\frac{3\gamma}{4\omega_1}I_{\rm ho}-\frac{1}{2\omega_1\sqrt{2I_{\rm ho}}}[F_1\cos\phi_{\rm ho}+ F_2\cos(\Delta\omega_2t-\phi_{\rm ho})],
\label{eq:phi}\\
\dot I_{\rm ho}&=-\frac{\partial\mathcal H_0}{\partial \phi_{\rm ho}}-\mathcal{F}_{I_{\rm ho}}\nonumber
\\ &=-\frac{\sqrt{2I_{\rm ho}}}{2\omega_1}[F_1\sin\phi_{\rm ho}- F_2\sin(\Delta\omega_2t-\phi_{\rm ho})]-2\Gamma I_{\rm ho}.
\label{eq:I}
\end{align}
For $\Gamma=0,F_2=0$, Eqs.~\eqref{Ham_AA}-\eqref{eq:I} can be readily integrated, and the action dynamics can be simply visualized as the motion of a "particle" trapped in a potential well of a particular shape. To see this, we simply square and add Eq.~\eqref{Ham_AA} and Eq.~\eqref{eq:I} to get
\begin{equation}
\frac{\dot I_{\rm ho}^2}{2}+U_{\rm eff}(I_{\rm ho})=0,
\end{equation}
where
\begin{align}
U_{\rm eff}(I_{\rm ho})=\frac{1}{2}\left(\mathcal{H}_0+\Delta\omega_1I_{\rm ho}-\frac{3\gamma I_{\rm ho}^2}{8\omega_1}\right)^2-\left(\frac{F_1}{2\omega_1}\right)^2I_{\rm ho}.
\label{eq:U_eff}
\end{align}
Thus, the particle with action $I_{\rm ho}$ oscillates in the effective potential $U_{\rm eff}$ with zero effective total energy. The time dependence of $I_{\rm ho}$ can be written explicitly in terms of the Jacobi elliptic fucntions as

\begin{align}
I_{\rm ho}=\frac{z_2I_1+z_1I_2-(z_2I_1-z_1I_2){\rm cn}(\tau-\tau_0)}{z_1+z_2-(z_1-z_2){\rm cn}(\tau-\tau_0)},
\label{eq:I(t)}
\end{align}
where $\tau=3\pi\gamma\sqrt{z_1z_2}t/[16\omega_1K(m_{\mathcal{H}_0})]$, $I_j$'s with $(j=1,2,3,4)$ are the roots of the equation $U_{\rm eff}(I_{\rm ho})=0$, $I_1>I_2$ are real, $I_4=I_3^*$ are complex conjugates, $z_j=\sqrt{(I_j-I_3)/(I_j-I_4)}$ with $(j=1,2)$, $K(m_{\mathcal{H}_0})$ is the complete elliptic integral of the first kind with the parameter $m_{\mathcal{H}_0}=[(I_1-I_2)^2-(z_1-z_2)^2]/(4z_1z_2)$, and $\tau_0$ is set by the initial condition.

For small non-zero $\Gamma\ll\Omega$ and $F_2\ll F_1$, the Hamiltonian $\mathcal{H}_0(I_{\rm ho},\phi_{\rm ho})$ is no longer a constant of motion. However, the evolution of $\mathcal{H}_0$ can be found from
\begin{equation}
\mathcal{\dot H}_0=-\left(\frac{\partial\mathcal{H}_0}{\partial\phi_{\rm ho}}\mathcal{F}_{\phi_{\rm ho}} +\frac{\partial\mathcal{H}_0}{\partial I_{\rm ho}}\mathcal{F}_{I_{\rm ho}}\right).\label{eq:H_S_ev}
\end{equation}

As the right-hand side (RHS) of Eq.~\eqref{eq:H_S_ev} is small (of order $\Gamma$ and $F_2$ in magnitude), its influence on $\mathcal{H}_0$ can be separated into two effects. The rapidly oscillating components cause only small-amplitude, high-frequency fluctuations(e.g., oscillate with the frequency $\Omega$). In contrast, the slowly varying components drive a persistent drift that accumulates into a large, long-term change. Therefore, to isolate the dominant long-term dynamics, we neglect these fast-oscillating terms on the RHS of Eq.~\eqref{eq:H_S_ev}. Neglecting the terms containing the fast oscillations ($\ee^{i n\Omega t}$, where $n = \pm 1,\pm2, ...$) is equivalent to averaging over the period of the oscillations $T = 2\pi/\Omega$\cite{Guckenheimer2013}; thus this method is often called the method of averaging. Hence, the averaged equation of $\mathcal{H}_0$ is given by
\begin{equation}
\langle\mathcal{\dot H}_0\rangle=-\left\langle\frac{\partial\mathcal{H}_0}{\partial\phi_{\rm ho}}\mathcal{F}_{\phi_{\rm ho}} +\frac{\partial\mathcal{H}_0}{\partial I_{\rm ho}}\mathcal{F}_{I_{\rm ho}}\right\rangle,\label{eq:H_S_ev_av}
\end{equation}
where $\langle f\rangle=T^{-1}\int_0^Tfdt$.

As explained in the main text and can be seen from Fig.~\ref{fig:S2}(c), when both $\Gamma$ and $F_2$ are non-zero, in steady-state they balance each other to produce a nearly constant $\langle\mathcal{H}_0\rangle$ for which $I_{\rm ho}$ oscillates periodically. As $F_2$ is increased, the value of $|\langle\mathcal{H}_0\rangle-\mathcal{H}_0(q_{\rm eq})|$ becomes larger and the amplitude of the attendant oscillations increases.

When $|[\mathcal{H}_0-\mathcal{H}_0(q_{\rm eq})]/\mathcal{H}_0(q_{\rm eq})|\ll1$, we find that $z_1\approx z_2$; thus, Eq. \eqref{eq:I(t)} can be approximated by $I_{\rm ho}\approx [I_1+I_2-(I_1-I_2)\cos(\Omega t+ \psi)]/2$, and Eq.~(2) of the main text by $q-q_{\rm eq}\approx q_1\cos(\Omega t+ \psi)+ q_2\cos(2\Omega t+ 2\psi),~p\approx p_1\sin(\Omega t+\psi).$ A straightforward perturbation analysis reveals that in the presence of small damping $(\Gamma\ll\Omega)$, $q_1$ and $p_1$ decay exponentially at the same rate, {\it i.e.}, $q_1,p_1\propto \ee^{-\Gamma t}$, while $q_2$ decays exponentially at twice the rate, {\it i.e.}, $q_2\propto \ee^{-2\Gamma t}$. Consequently, when the lightly damped system is prepared with a fixed value of $\mathcal{H}_0$ that corresponds to fixed $I_1$ and $I_2$ such that $|[\mathcal{H}_0-\mathcal{H}_0(q_{\rm eq})]/\mathcal{H}_0(q_{\rm eq})|\ll1$, we calculate the ratio $q_2(0)/q_1(0)=[3\gamma(I_1-I_2)/(32\omega_1)][\delta\omega_1-3\gamma(I_1+I_2)/(8\omega_1)]^{-1}$. Thus, the decay of $q(t)$ towards $q_{\rm eq}$ is overall non-exponential and is dominated by a decay rate of $2\Gamma$ $(|q|\propto \ee^{-2\Gamma t})$ when $[3\gamma(I_1-I_2)/(32\omega_1)][\delta\omega_1-3\gamma(I_1+I_2)/(8\omega_1)]^{-1}\gg1$ and by a decay rate of $\Gamma$ $(|q|\propto\ee^{-\Gamma t})$ when $[3\gamma(I_1-I_2)/(32\omega_1)][\delta\omega_1-3\gamma(I_1+I_2)/(8\omega_1)]^{-1}\ll1$.

\section{Phase-Space Trajectories in the Rotating Frame}

\begin{figure}[H]
\centering
\includegraphics[width=0.8\linewidth]{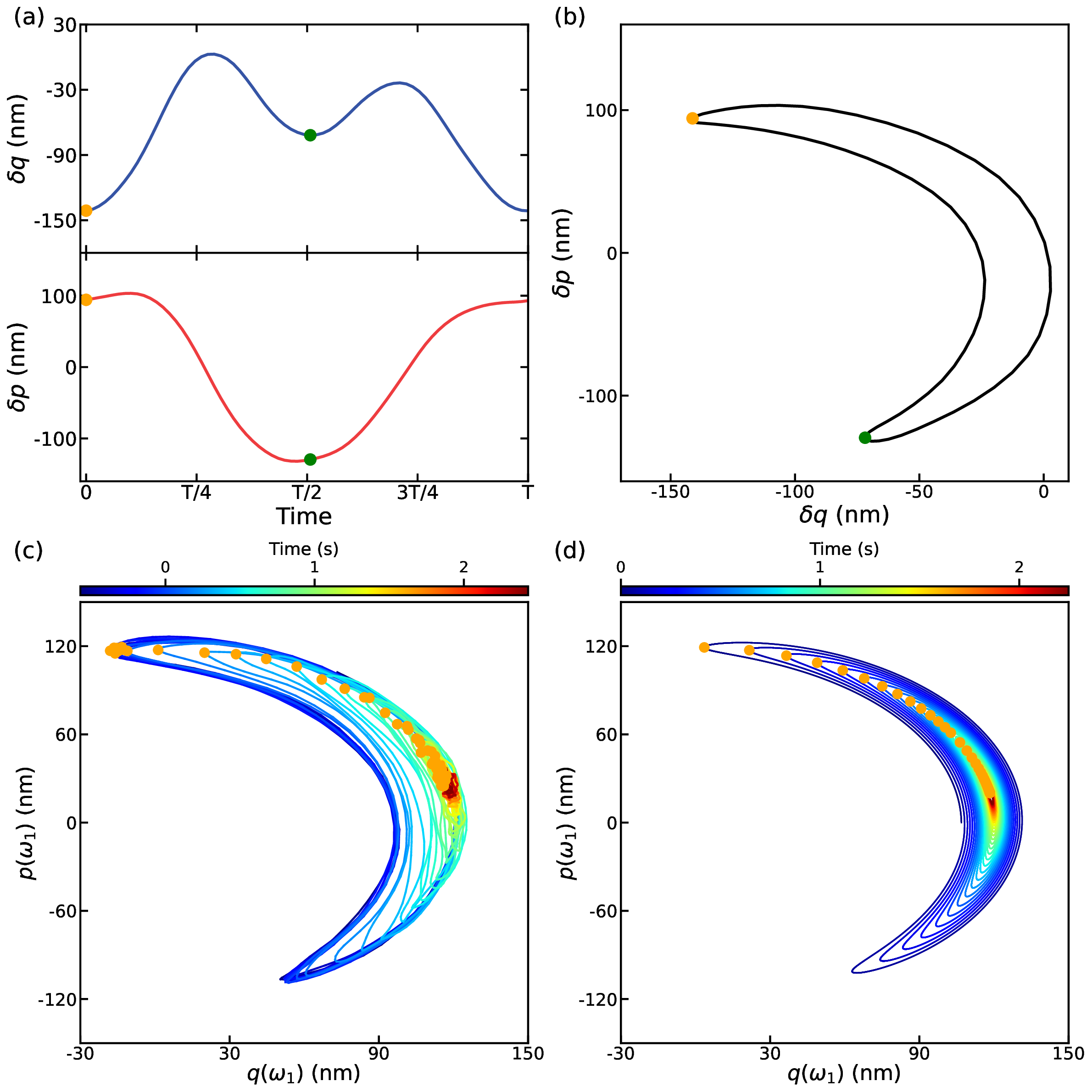}
\caption{(a) The oscillation of $\delta q$ and $\delta p$ within one cycle of the first harmonic frequency $T_2 = 2\pi/\delta \omega_2$, \( \delta \omega_2/2\pi = 13 \) Hz and \( F_2 = \qty{150}{\milli \meter/ \second^2} \), with the same values of ($F_1$, $\delta\omega_1$) used for Fig.~3 of the main text. Orange and green circles indicate the extremal points. (b) Phase space trajectories in the rotating frame for the same cycle of (a) with the same extremal points indicated. Note that the orange circle represents the maximal deviation from equilibrium. (c),(d) Phase space trajectories in the rotating frame over time during the ring-down process, reconstructed from Fig.~3(a) of the main text, with the colormap representing the progression of time. Panels (c) and (d) show experimental and theoretical trajectories, respectively. The orange circles denote the points of maximal deviation from equilibrium in each cycle of the trajectory.}
\label{fig:S1}
\end{figure}

The points of maximal deviation from equilibrium in the rotating frame phase space correspond to the lower and upper envelopes of the quadratures $q$ and $p$, as illustrated in Fig.~3(a) of the main text and Fig.~\ref{fig:S1}. The term ''maximum'' refers to a point where the first derivative of $|\delta q|$ is zero, the second derivative is negative, and, among the two such extrema in the upper panel of Fig.~\ref{fig:S1}(a), the larger one is chosen. The attendant value of $p$ is shown in the lower panel of Fig.~\ref{fig:S1}(a). These maximal deviation points serve as reference markers for analyzing the decay behaviors of $q$ and $p$ during the ring-down process.

Another straightforward approach to analyzing the ring-down signal is to compare the experimental phase-space trajectories of ($q$, $p$) in the rotating frame with their theoretical counterparts and observe how these trajectories decay over time. The time evolution of the area \( \mathcal{S} \), which is enclosed by the nearly-closed trajectories and is proportional to the action, can be obtained by using the divergence theorem to show that
\begin{equation}
\dot{I}\propto\dot{\mathcal{S}} = \oint_\ell \bm{f} \cdot \bm{n} \, d\ell
=\int_{\mathcal{S}}\nabla\cdot\bm{f}d{\mathcal{S}}= -2 \Gamma \mathcal{S}. \label{eq:path_int}
\end{equation}
In this expression, $I=(2\pi)^{-1}\oint p dq$ is the action associated with $\mathcal{H}_0$, \( \bm{f} = (\dot{q}, \dot{p})^T \) represents the velocity vector in the phase portrait, \( \bm{n} \) is the normal vector to the trajectory, and \( d\ell \) is an infinitesimal segment of the closed trajectory. From Eq.~\ref{eq:path_int} we find that $\mathcal{S}(t)=\mathcal{S}(0)e^{-2\Gamma t}$.

After switching off $F_2$, the area $\mathcal{S}$ decays as the system’s trajectories contract within the rotating frame as shown in Fig.~\ref{fig:S1}(c), (d). The logarithmic plot of $\mathcal{S}$ over time in Fig.~3(c) of the main text shows good agreement with the theoretical prediction, exhibiting a slope of $2\Gamma\approx\qty{2.7}{\hertz}$.

\section{Experimental Method and Quasi-Conservative Dynamics}

By applying an AC current in the presence of a perpendicular magnetic field, as depicted in Fig.~1(a) of the main text, the device is harmonically driven primarily via the Lorentz force and generates the electromotive force \cite{Cleland1996}. This signal is fed into a Zurich Instruments HF2LI lock-in amplifier, which demodulates the in-phase and quadrature components. A secondary drive is introduced to excite the sidebands by adding it to the output source of the lock-in amplifier. The sideband signal in the linear regime is measured using a second demodulator. In the nonlinear regime, a homodyne measurement was carried out to encapsulate the more complicated full dynamics of the system.

\begin{figure}[t]
\centering
\includegraphics[width=1\linewidth]{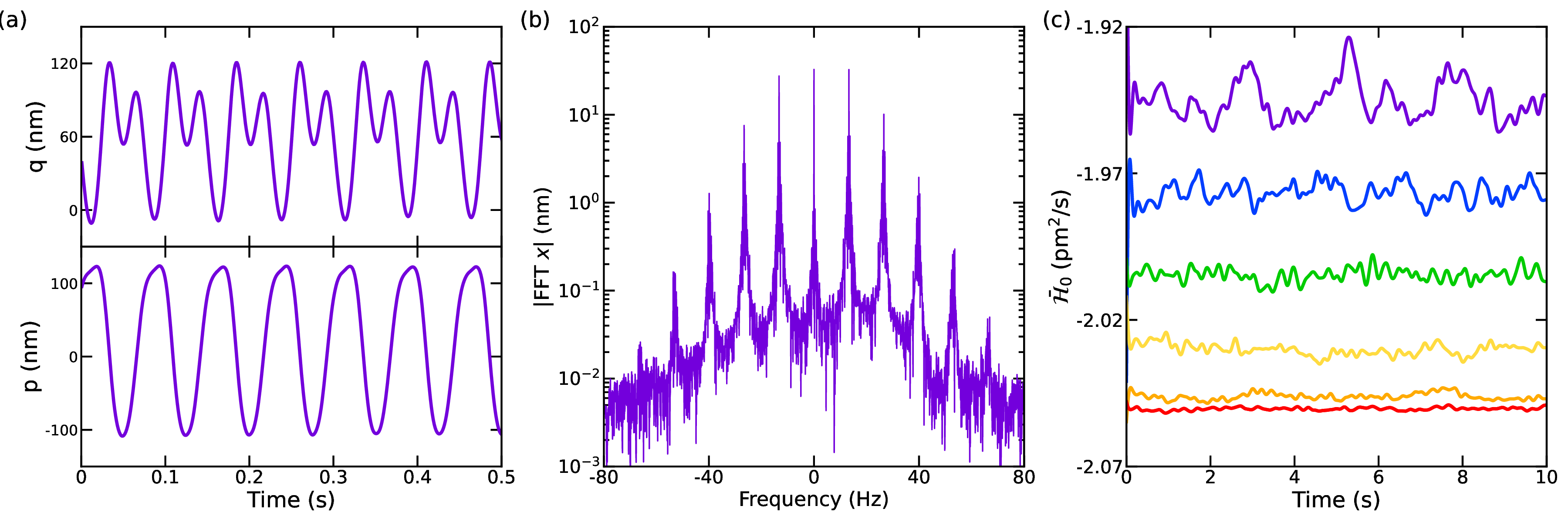}
\caption{(a) The top and bottom panels show the in-phase and quadrature components over time at the peak condition with $\delta\omega_2=13.3$ Hz and $F_2=\qty{150}{\milli \meter/\second^2}$, corresponding to the purple-colored case in Fig.2 and Fig.4. of the main text. (b) The measured spectrum of $x(t)=q(t)\cos(\omega_1 t)+p(t)\sin(\omega_1 t)$ in the laboratory frame, using the $q(t)$ and $p(t)$ components shown in the top and bottom panels in (a). (c) The filtered Hamiltonian $ \mathcal{\bar{H}}_0 $ for each drive level of $F_2$, represented using the same rainbow colors in the main text, low-pass filtered with a cutoff frequency of 5 Hz. Drive levels are $F_2=15,~30,~60,~90,~120,$ and 150 $\rm m m/s^2$.}
\label{fig:S2}
\end{figure}

Strong nonlinearity has far-reaching consequences on the relaxation of the quadratures, which can be detected via homodyne measurement of $q(t)$ and $p(t)$ as shown in Fig.~\ref{fig:S2}(a). In particular, the dynamics of $p(t)$ is dominated by its fundamental harmonic $p\propto\exp(\pm i\delta\omega_2t)$ as shown in Fig.~4(d) of the main text. As a result, the overall spectrum of $x(t)$ in Fig.~\ref{fig:S2}(b) are masked by considering only $p$.

An increase in the peak amplitude at $\omega_2$ with increasing $F_2$ corresponds to higher-energy states in the rotating frame as captured by the system's low-pass filtered Hamiltonian level as shown in Fig.~\ref{fig:S2}(c). The time-averaged Hamiltonian defined as $\langle \mathcal{H}_0\rangle=T_2^{-1}\int_{0}^{T_2}\mathcal{H}_0dt$, serves as the horizontal axis variable for Fig.~4(c), (d) in the main text. The low-pass filtered Hamiltonian remains approximately constant at each drive level, shown in Fig.~\ref{fig:S2}(c), indicating a quasi-conservative system due to the sideband damping being compensated by the secondary drive $F_2$.

\section{Frequency-energy relation}

Fig.~2(a) of the main text shows the results of a frequency sweep of the upper sideband in which $\omega_2$ is varied, resulting in different energy states in the rotating frame. The figure shows only the amplitude of the first harmonic.
However, because the system is strongly nonlinear, multiple harmonics must be considered to accurately capture the nonlinear response. To show this, we introduce the frequency-energy relation based on the conservative Hamiltonian dynamics, neglecting damping so that the trajectories remain centered at $(q_{\rm eq}, 0)$. We observe that the effective resonance frequency in the rotating frame monotonically decreases as $\mathcal{H}_0$ increases, as depicted in Fig.~S3(a). In the rotating frame phase space, the trajectories correspond to increasingly large, banana-shaped loops (Fig.~S3(b)). Furthermore, higher-order harmonics of $q$ become significant as $\mathcal{H}_0$ increases. Over the $\mathcal{H}_0$ values of interest the first and second harmonics dominate the response. Since these components have slower decay rates than the higher harmonics, they are essential for analyzing the ring-down behavior shown in Fig.~3 of the main text.

\begin{figure}[t]
\centering
\includegraphics[width=1\linewidth]{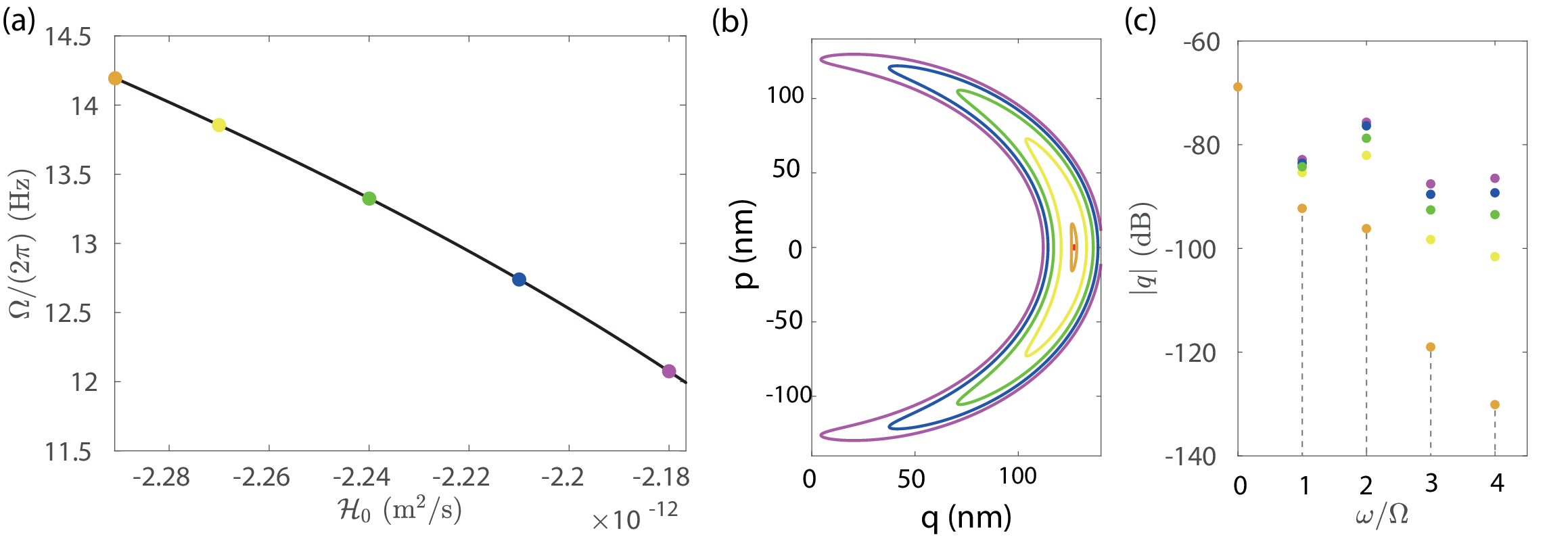}
\caption{ (a) Effective resonance frequency in the rotating frame as a function of the conservative Hamiltonian $\mathcal{H}_0$. The Black line indicates the theoretical prediction, and the rainbow-colored circles denote representative points along the curve. (b) Theoretical phase-space trajectories in the rotating frame corresponding to the representative points in (a). (c) Theoretical harmonic components of $q$, plotted using the same color scheme as in (a)}
\label{fig:S3}
\end{figure}

\section{Ring-down Measurement}

In the ring-down measurement presented in Fig.~1(c) of the main text, the system was initially driven in the Duffing nonlinear regime. After switching off the drive $F_1$, the demodulated signal at $\omega_0$ was monitored using a sufficiently large bandwidth ($>\delta \omega_1$) to capture the full ring-down dynamics originating from the nonlinear state\cite{maillet2016}.

In the ring-down measurement of the rotating frame, there exist two distinct time scales: the fast dynamics of the main tone at $\sim\omega_1^{-1}$ and the slower oscillatory motion at $\sim\delta\omega_2^{-1}$ and decay at $\sim\Gamma^{-1}$. Our real-time measurement technique is designed to track the slow dynamics of the motion, which oscillates at a frequency of $\delta\omega_2$.

To achieve this, we measure the real-time quadrature components $q(t)$ and $p(t)$, which are demodulated using reference signals $\cos(\omega_1 t)$ and $\sin(\omega_1 t)$. The key requirement is that the demodulator’s bandwidth at $\omega_1$ must be sufficiently broad to capture the full harmonic content of the slow dynamics, preventing any unintentional filtering of relevant signals\cite{Zurich_lock_in}. In terms of time constant, which is inversely related to its bandwidth, the demodulator's time constant must be small enough to resolve the slow oscillations, {\it i.e.}, smaller than $\delta\omega_2^{-1}$.

For ring-down measurements, an additional constraint arises from the dissipation rate $\Gamma$. The demodulator's time constant must be shorter than the characteristic dissipation time $(2\Gamma)^{-1}$ of the oscillation. If the time constant is too large ($>(2\Gamma)^{-1}$), the demodulated signal becomes excessively smoothed, preventing accurate observation of the ringdown behavior.

As a concrete example in our system, for $\delta \omega_2/2\pi \sim 10$ Hz, we set the demodulator’s time constant to 2 ms. This value is sufficiently small compared to both $(2\Gamma)^{-1} \sim 370$ ms and $\delta\omega_2^{-1}\sim10$ ms, ensuring that our real-time measurement method accurately captures the oscillatory motion and its decay dynamics.

\section{Qualitative Understanding of Strong Nonlinearity}

To qualitatively understand the strong second harmonic of $q$, we decompose the effective Hamiltonian $\mathcal{H}_0$ into three parts: harmonic component \( \mathcal{H}_{L} = \delta \omega_1 (q^2 + p^2 )/2 \), symmetric nonlinear component $\mathcal{H}_{NL}=3\gamma(q^2+p^2)^2/32\omega_1$, and asymmetric component $\mathcal{H}_{A}=-F_{1}q/2\omega_1$. The coefficients \( \delta \omega_1 \) and $F_1$ are key tunable parameters of the system's nonlinearity in the rotating frame, revealing a competing relationship between harmonicity (symmetry) and anharmonicity (asymmetry). When $\delta \omega_1/F_1$ is large—proportional to $\mathcal{H}_L/\mathcal{H}_A$—the system remains primarily harmonic, suppressing the effects of anharmonicity. Conversely, when the ratio $\delta \omega_1/F_1$ is small, the anharmonic potential becomes more prominent.

In Fig.~2(b) of the main text and Fig.~S4(a), the estimated values at equilibrium ($q_\text{eq}$, $p_{\rm eq}$) are $\mathcal{H}_{L}\sim\qty{-d-12}{\meter^{2}\hertz}$ and $\mathcal{H}_{A}\sim\qty{-d-13}{\meter^{2}\hertz}$, which are relatively comparable. As a result, when the system is driven sufficiently hard by $F_2$, the dynamics enter a nonlinear regime where $\mathcal{H}_A$, which depends solely on $q$, plays a key role in the emergence of the strong second harmonic of $q$. In contrast, for Fig.~S4(b) where $\delta \omega_1/F_1$ is relatively large, these values shift to $\mathcal{H}_{L}\sim-10^{-7} \; \rm{m^2 /s}$ and $\mathcal{H}_{A}\sim-10^{-12} \; \rm{m^2 /s}$, making the anharmonic effects much weaker.

\begin{figure}[t]
\centering
\includegraphics[width=0.8\linewidth]{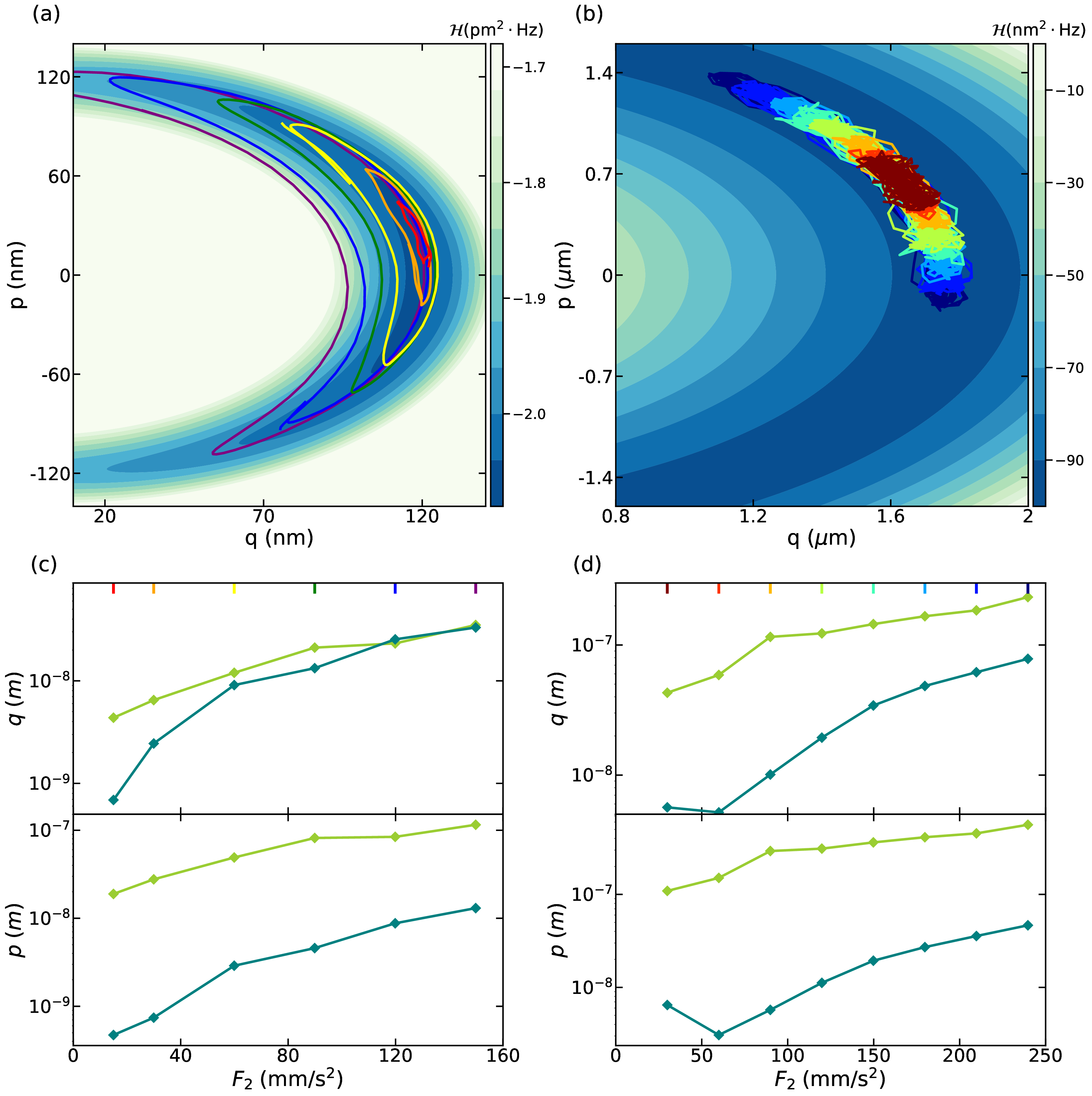}
\caption{Comparison of nonlinearities for small and large values of $\delta \omega_1/F_1$. (a), (c) correspond to the small case (\(\delta \omega_1 = 86.9\) Hz, \( F_1 = 3.0 \, \mathrm{m/s^{2}} \)), while (b), (d) correspond to the large case (\(\delta \omega_1 = 18.4\) kHz, \( F_1 = 60 \, \mathrm{m/s^{2}} \)). (a),(b) The phase space representation of the Hamiltonian $\mathcal{H}_0$ in the rotating frame near the equilibrium. Its contours represent constant Hamiltonian levels from the model, color-mapped accordingly. The maximum Hamiltonian levels of (a),(b) are capped differently for clear visualization of the region of interest. Rainbow-colored lines denote experimentally obtained trajectories at the peak condition by varying \( F_2 \) and $\delta \omega_2$. $F_2$ values are indicated on the x-axes of (c),(d), with the corresponding values marked on the upper x-axis of the top panels. (c),(d) First and second harmonic components of \( q \) and \( p \) as a function of \( F_2 \).}
\label{fig:S4}
\end{figure}

The secondary drive \( F_2 \) ranges from \qty{15}{\milli \meter/ \second^{2}} to \qty{150}{\milli \meter / \second^{2}} for \( \delta \omega_1 = \qty{86.9}{\hertz}\), $F_1= \qty{3.0}{\meter/ \second^{2}}$ in Fig.~\ref{fig:S4}(a), and from \qty{30}{\milli \meter /\second^{2}} to \qty{240}{\milli \meter /\second^{2}} for \( \delta \omega_1 = \qty{18.4}{\kilo \hertz}\), $F_1= \qty{60}{\meter/ \second^{2}}$ in Fig.~\ref{fig:S4} (b). For each $F_2$, the detuning of the secondary drive, $\delta \omega_2$, is chosen to correspond to the peak values (Fig.~2(a) in the main text). For small ratio $\delta \omega_1/F_1$ at equilibrium, the asymmetric term becomes comparable to the linear, leading to asymmetric trajectories in the phase space of the rotating frame. This results in a noticeable comparable contribution from the first and second harmonics, with the second harmonic of $q$ becoming increasingly prominent (the upper panel of Fig.~\ref{fig:S4} (c)). For large detuning, the anharmonic term is relatively small, resulting in a more symmetric shape (Fig.~\ref{fig:S4}(b)) and a weaker second-harmonic component of $q$ (the upper panel of Fig.~\ref{fig:S4} (d)). This comparison explains why, in the case of Fig.~\ref{fig:S4}(a), strong nonlinear behavior can be achieved with a relatively small secondary drive $F_2$, as the harmonic and asymmetric contributions remain comparable.

\section{Nonlinear Damping effect}
\begin{figure}[H]
\centering
\includegraphics[width=0.5\linewidth]{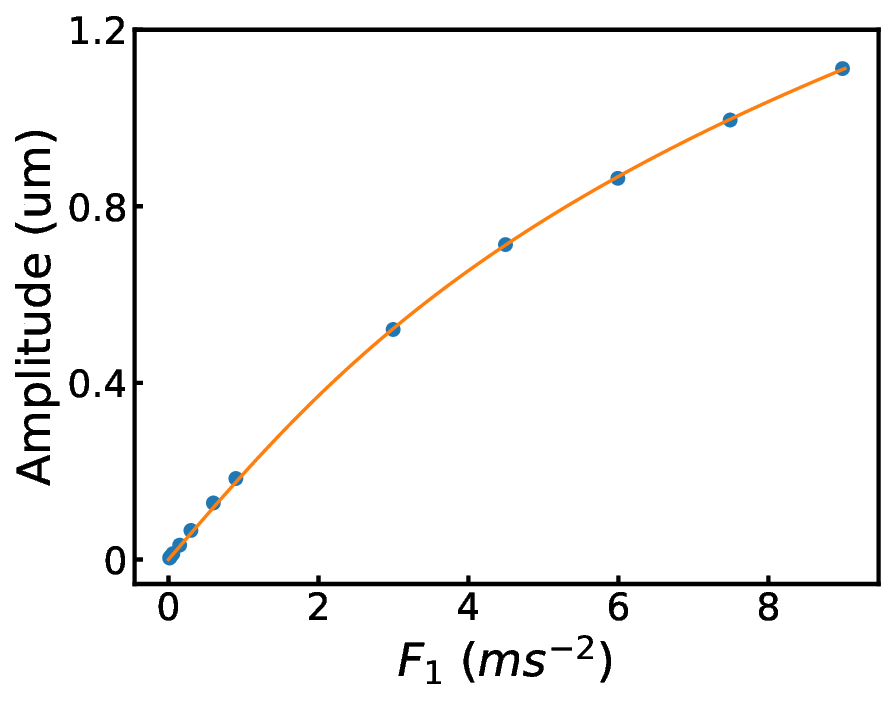}
\caption{Peak amplitude as function of $F_1$. The solid line represent the fitting line.}
\label{fig:S5}
\end{figure}

Although the fundamental flexural mode exhibits some level of nonlinear damping in the large amplitude regime, the experimental condition in the main text ensures that its effect is negligible. From the fitting in Fig.~\ref{fig:S5}, the estimated nonlinear damping coefficient is $\Gamma_2\simeq \qty{6.34 d12}{\meter^{-2}\second^{-1}}$ and $\Gamma_{2}q_{\text{eq}}^{2}\simeq\qty{0.09}{\hertz}$, which is sufficiently small compared to the linear damping $\Gamma_2q_{\text{eq}}^2/\Gamma_1 \ll1$\cite{nano_tune}.

\bibliography{references}